%% file: ms.tex
\documentclass[twocolumn,tighten,times,astrosymb,floatfix]{aastex631}

\usepackage{
	amsmath,
	amssymb,
	amsthm,
	booktabs,
	epsfig,
	graphicx,
	grffile,
	import,
	morefloats,
	microtype,
	natbib,
	subfigure,
	ulem,
	url,
	verbatim,
	wasysym,
	xcolor,
	xspace,
}
\usepackage{savesym}
\savesymbol{tablenum}
\usepackage{siunitx}
\restoresymbol{SIX}{tablenum}
\usepackage[version=4]{mhchem}

\input{commands.tex}

\shorttitle{MRS Beta Pic}
\shortauthors{Worthen et al.}

\begin{document}

\title{\Large
 MIRI MRS Observations of Beta Pictoris I. The Inner Dust, the Planet, and the Gas
}

\correspondingauthor{Kadin Worthen}
\email{kworthe1@jhu.edu}

\author[0000-0002-5885-5779]{Kadin Worthen}
\affiliation{William H. Miller III Department of Physics and Astronomy, John's Hopkins University, 3400 N. Charles Street, Baltimore, MD 21218, USA}

\author[0000-0002-8382-0447]{Christine H. Chen}
\affiliation{Space Telescope Science Institute, 3700 San Martin Drive, Baltimore, MD 21218, USA }
\affiliation{William H. Miller III Department of Physics and Astronomy, John's Hopkins University, 3400 N. Charles Street, Baltimore, MD 21218, USA}

\author[0000-0002-9402-186X]{David R. Law}
\affiliation{Space Telescope Science Institute, 3700 San Martin Drive, Baltimore, MD 21218, USA }

\author[0000-0001-9352-0248]{Cicero X. Lu}
\affiliation{Gemini Observatory/NSF’s NOIRLab, 670N. A’ohoku Place, Hilo, HI 96720, USA}

\author[0000-0002-9803-8255]{Kielan Hoch}
\affiliation{Space Telescope Science Institute, 3700 San Martin Drive, Baltimore, MD 21218, USA }

\author{Yiwei Chai}
\affiliation{William H. Miller III Department of Physics and Astronomy, John's Hopkins University, 3400 N. Charles Street, Baltimore, MD 21218, USA}

\author[0000-0003-4520-1044]{G.C. Sloan}
\affiliation{Space Telescope Science Institute, 3700 San Martin Drive, Baltimore, MD 21218, USA }
\affiliation{Department of Physics and Astronomy, University of North Carolina, Chapel
  Hill, NC 27599-3255, USA }

\author[0000-0001-9855-8261]{B. A. Sargent}
\affiliation{Space Telescope Science Institute, 3700 San Martin Drive, Baltimore, MD 21218, USA }
\affiliation{William H. Miller III Department of Physics and Astronomy, John's Hopkins University, 3400 N. Charles Street, Baltimore, MD 21218, USA}

\author[0000-0003-2769-0438]{Jens Kammerer}
\affiliation{European Southern Observatory, Karl-Schwarzschild-Straße 2, 85748 Garching, Germany}
\affiliation{Space Telescope Science Institute, 3700 San Martin Drive, Baltimore, MD 21218, USA }

\author[0000-0003-4653-6161]{Dean C. Hines}
\affiliation{Space Telescope Science Institute, 3700 San Martin Drive, Baltimore, MD 21218, USA }

\author[0000-0002-4388-6417]{Isabel Rebollido}
\affiliation{Centro de Astrobiología (CAB), INTA-CSIC, Camino Bajo del Castillo s/n - Villafranca del Castillo, 28692 Villanueva de la Cañada, Madrid, Spain }

\author[0000-0001-6396-8439]{William Balmer}
\affiliation{William H. Miller III Department of Physics and Astronomy, John's Hopkins University, 3400 N. Charles Street, Baltimore, MD 21218, USA}

\author[0000-0002-3191-8151]{Marshall D. Perrin}
\affiliation{Space Telescope Science Institute, 3700 San Martin Drive, Baltimore, MD 21218, USA }

\author[0000-0001-8302-0530]{Dan M. Watson}
\affiliation{Department of Physics and Astronomy, University of Rochester, 500 Wilson Blvd, Rochester, NY 14627, USA}

\author[0000-0003-3818-408X]{Laurent Pueyo}
\affiliation{Space Telescope Science Institute, 3700 San Martin Drive, Baltimore, MD 21218, USA }

\author[0000-0001-8627-0404]{Julien H. Girard}
\affiliation{Space Telescope Science Institute, 3700 San Martin Drive, Baltimore, MD 21218, USA }

\author[0000-0002-9548-1526]{Carey M. Lisse}
\affiliation{Johns Hopkins University Applied Physics Laboratory, 11100 Johns Hopkins Rd, Laurel, MD 20723, USA}

\author{Christopher C. Stark}
\affiliation{NASA Goddard Space Flight Center, Exoplanets $\&$ Stellar Astrophysics Laboratory, Code 667, Greenbelt, MD 20771, USA}

\begin{abstract}
 We present JWST MIRI Medium Resolution Spectrograph (MRS) observations of the $\beta$ Pictoris system. We detect an infrared excess from the central unresolved point source from 5 to 7.5 $\text{\textmu}$m which is indicative of dust within the inner $\sim$7 au of the system. We perform PSF subtraction on the MRS data cubes and detect a spatially resolved dust population emitting at 5 $\text{\textmu}$m. This spatially resolved hot dust population is best explained if the dust grains are in the small grain limit (2$\pi$a$\ll$$\lambda$). The combination of unresolved and resolved dust at 5 $\text{\textmu}$m could suggest that dust grains are being produced in the inner few au of the system and are then radiatively driven outwards, where the particles could accrete onto the known planets in the system $\beta$ Pic b and c.  We also report the detection of an emission line at 6.986 $\text{\textmu}$m that we attribute to be [Ar II]. We find that the [Ar II] emission is spatially resolved with JWST and appears to be aligned with the dust disk. Through PSF subtraction techniques, we detect $\beta$ Pic b at the 5$\sigma$ level in our MRS data cubes and present the first mid-IR spectrum of the planet from 5 to 7 $\text{\textmu}$m. The planet's spectrum is consistent with having absorption from water vapor between 5 and 6.5 $\text{\textmu}$m. We perform atmosphere model grid fitting on spectra and photometry of $\beta$ Pic b and find that the planet's atmosphere likely has a sub-stellar C/O ratio.  
\end{abstract}



\section{Introduction}\label{sec:Introduction}

Debris disks are planetary systems that consist of dust, gas, planetesimals, and planets that typically correspond to the late stages of planetary system formation \citep{wyatt08,Hughes18}. They provide a unique laboratory to study the processes involved in the later stages of planet formation and evolution. Unlike protoplanetary disks, the dust seen in debris disks is thought to be constantly replenished through collisional processes between minor bodies in the system \citep{Hughes18}. This is because the dust grains in debris disks are subject to radiation pressure and Poynting-Robertson drag that remove dust from orbits around the central star on timescales that are short compared to the age of the system \citep{Guess62,Krivov06,wyatt2008}. The detection of these dust grains in debris disks points to ongoing stochastic and stead-state collisions between planetesimals that actively replenish the small dust grains \citep{Backman93}. 

The particles in debris disks range in size from the parent planetesimals to the collisionally produced dust, but it is the dust that is observable in both thermal emission at mid-infrared to millimeter wavelengths (e.g., \citealt{koerner98,Holland98,Telesco05,macgregor18}) and scattered light at optical and near-infrared wavelengths (e.g., \citealt{Smith84,Kalas06,Esposito20,ren23}). Infrared spectra and SEDs of debris disks can reveal the temperature of the dust (e.g., \citealt{ballering13}), which depends on the location, size, and composition of the dust particles. SED modeling as well as multi-wavelength imaging have revealed some debris disks to have multiple populations of dust at various stellocentric distances (e.g., \citealt{Chen14,Condell15,Gaspar23}). Knowledge of the inner most regions of debris disks has been mostly limited to infrared spectroscopic and photometric analyses, mainly because of the low spatial resolution of previous space-based infrared observatories like Spitzer. JWST provides a unique opportunity to study the spectra and structure of dust in debris disks at high angular resolution with the MIRI Medium Resolution Spectrograph (MRS).  

$\beta$ Pictoris is a $\sim$23 Myr \citep{Mamajek14} A6V star that is host to the first ever imaged debris disk \citep{Smith84}. The $\beta$ Pic disk is oriented close to edge-on from our line of sight and has been studied in scattered light as well as thermal emission at mid-infrared and millimeter wavelengths (e.g., \citealt{Telesco05,Golimowski06,Matra19,Rebollido23}). At a distance of 19.6 pc \citep{GAIA23}, $\beta$ Pic provides a great laboratory for studying the spatial structure of its debris disk.
Scattered light imaging with HST revealed a warp in the inner disk \citep{Golimowski06} that was later attributed to interactions with the now confirmed giant planet $\beta$ Pic b ($\sim$9 M$_\text{J}$, $a=$9.9 au) \citep{Lagrange09,Dawson11,Lagrange12,Nowak20,2Nowak20}. Radial velocities provided evidence for the presence of a second planet in the system, $\beta$ Pic c \citep{Lagrange19} ($\sim$8 M$_\text{J}$, $a=$2.7 au), which was later directly confirmed by interferometric observations with VLT/GRAVITY \citep{2Nowak20}. 

As a young nearby system with both giant planets and a debris disk, $\beta$ Pic provides a unique opportunity to study the interactions between dust in the disk and the giant planets in the system. Giant planets present in debris disks are thought to impart structures on the dust in debris disks (e.g., \citealt{Dawson11,Crotts21}), however, it is unclear if and to what extent the dust and minor bodies in debris disks can affect giant planets via accretion. Giant planets likely form within $\sim$10 Myr, before the debris disk evolutionary stage \citep{Williams11,Li16}, although it is still possible for material from debris disks to accrete onto giant planets (e.g., \citealt{Kral20,Frantseva22}). In our Solar System for instance, the impact of comet Shoemaker-Levy 9 in 1994 delivered refractory material to Jupiter's atmosphere \citep{Harrington04,Fletcher10}. \cite{Marley12} suggested that micron and sub-micron sized silicate grains can remain at higher altitudes in the atmospheres of young, low surface gravity giant planets and brown dwarfs and thus affect their observed spectra. If young giant planets are accreting dust from their debris disks, it is then possible for the dust grains to remain in their atmosphere and affect the observed spectrum of the planet. For young giant planets that are in systems with debris disks, the amount of dust that is accreted from the disks onto the planets is not well observationally constrained.   

Mid-infrared spectroscopy of $\beta$ Pic has revealed the presence of small sub-micron silicate grains in the system \citep{Okamoto04,Li12,Lu22}. By modeling silicate features at 10, 18, and 23 $\text{\textmu}$m detected with {Spitzer IRS, \cite{Lu22} found that the features are best produced by dust grains in the Rayleigh limit (2$\pi$a$\ll$$\lambda$), indicating the presence of sub-micron sized silicate grains that are subject to blowout from radiation forces. Depending on the location of the dust, these sub-blowout sized grains have the potential to interact with the known planets in the system as they are being radiatively driven outwards. \cite{Lu22} also found tentative evidence for an infrared excess at 5 $\text{\textmu}$m and the presence of $\sim600$ K dust in the system. Because of the limited angular resolution of Spitzer, the location of this hot dust was not clear.

In this paper, we present JWST MIRI MRS observations of the $\beta$ Pic system which provide higher spatial and spectral resolution space-based data in the mid-IR than previously obtained with Spitzer. The observations and data reduction are described in Section 2. In Section 3, we present our main findings including (1) an infrared excess from 5 to 7.5 $\text{\textmu}$m, (2) the discovery of a spatially resolved hot dust population, (3) the first detection of [Ar II] at 6.986 $\text{\textmu}$m in the $\beta$ Pic disk and (4) the extraction of a low resolution MRS spectrum of $\beta$ Pic b. In section 4, we discuss the implications of the new spatially resolved hot dust population and how, along with the 5 $\text{\textmu}$m excess, it provides evidence for an outflowing wind of small dust grains that could be blown onto the known planets where they may be accreted. In section 5, we state our conclusions and summarize.

\section{Observations and Data Processing}\label{sec:Observations}
\subsection{Data Acquisition}
As a part of GTO program 1294, we observed $\beta$ Pictoris (K=3.48, \citealt{ducati02}) with MIRI MRS \citep{Wells15,Argyriou23} on 2023, January 11, preceded by dedicated background observations and followed immediately by observations of the nearby star N Car (A0II, K=4.218,\citealt{houk78,Cutri03}), which we used as a PSF reference star. N Car is offset from $\beta$ Pic by 7.6 degrees. All observations were taken in all three grating settings (short, medium, long) in all four channels with the FASTR1 readout pattern to cover a wavelength range of 4.9-28 $\text{\textmu}$m. We used a 4-point point-source dither pattern with the negative dither orientation\footnote{\url{https://jwst-docs.stsci.edu/jwst-mid-infrared-instrument/miri-operations/miri-dithering/miri-mrs-psf-and-dithering}} to observe $\beta$ Pic and N Car. We observed both stars with target acquisition to ensure that both stars were well aligned on the detector to optimize reference PSF subtraction. Both $\beta$ Pic and N Car were used themselves as target acquisition stars. The aperture position angle of JWST was 23.8 degrees for the observations of $\beta$ Pic to make sure that the MRS image slicers were orthogonal to the edge-on disk. This was to ensure that the MRS cross artifact \citep{Argyriou23} would not mimic disk signatures in the IFU data cubes. 

For the dedicated background observations, we used a 2-point dither pattern. The position of the dedicated background field was RA: 05 47 6.3034, Dec: -51 02 55.85, which is 1.7 arcminutes offset from $\beta$ Pic. The exposure time for N Car was longer than for $\beta$ Pic to ensure that the N Car observations had a similar S/N as the $\beta$ Pic observations. For $\beta$ Pic, the number of groups and integrations for each dither position for channels 1 and 2 (4.9-11.71 $\text{\textmu}$m) was 5 and 14 respectively (exposure time = 921 s). The number of groups and integrations for each dither position for channels 3 and 4 (11.55-28.1 $\text{\textmu}$m) was 15 and 5 (exposure time = 877 s). For N Car, the number of groups and integrations for each dither position for channels 1 and 2 was 5 and 28 (exposure time = 1854 s). The number of groups and integrations for each dither position for channels 3 and 4 was 15 and 10 (exposure time = 1765 s). For the dedicated background observation, the number of groups and integrations for each dither position for  channels 1 and 2 was 5 and 16 (exposure time = 264 s). For channels 3 and 4, the number of groups and integrations for the background observations was 15 and 6 (exposure time = 264 s).\footnote{All of the data presented in this article were obtained from the Mikulski Archive for Space Telescopes (MAST) at the Space Telescope Science Institute. The specific observations analyzed can be accessed via \dataset[DOI]{https://archive.stsci.edu/doi/resolve/resolve.html?doi=10.17909/7xb7-hh14}. }

\subsection{Data Reduction}
We processed the raw detector files for $\beta$ Pic and N Car through version 1.11.0 of the JWST calibration pipeline \citep{Bushouse23} using CRDS (Calibrated Reference Data System) context ``jwst$\_$1094.pmap" . The pipeline consists of 3 stages: \texttt{Detector1}, \texttt{Spec2}, and \texttt{Spec3}. We processed the $\beta$ Pic and N Car raw files with the exact same pipeline setup.

The \texttt{Detector1} stage converts the raw ramp images to uncalibrated slope images and also includes a jump detection step that flags jumps in the ramp between consecutive groups. This step mitigates ramp jumps that are often caused by cosmic ray hits. We changed the three group rejection threshold to be 100$\sigma$ in the jump detection step because the default setting in the pipeline over-flags jumps in the raw data and creates artifacts in the final spectrum. The default three group rejection threshold in the pipeline was 6$\sigma$. The \texttt{Spec2} pipeline takes the uncalibrated slope images and applies calibrations and instrumental corrections including a fringe correction. The MRS is known to suffer from fringing that can have effects of up to 30$\%$ of the spectral baseline (e.g., \citealt{argyriou20}). In the \texttt{Spec2} pipeline, we applied both fringe corrections available in the pipeline, the fringe flat and the residual fringe correction step, which is not turned on by default. In \texttt{Spec2}, we also performed the stray light subtraction step, which is necessary to subtract the cross-artifact.  

We used the \texttt{Spec3} pipeline to combine all 4 dither positions into spectral cubes for each of the 12 sub-bands. We created the spectral cubes with the \texttt{IFUalign} coordinate system so that the data cubes were aligned with the MRS image slicers and the PSFs of $\beta$ Pic and N Car were well aligned for optimal PSF subtraction. The background subtraction from the dedicated background observations was done in the \texttt{Spec3} pipeline along with the outlier rejection step. We built the spectral cubes using the \texttt{drizzle} algorithm in the pipeline \citep{Law23}. We left the pixel size to be the pipeline default which is 0.13'' for channel 1, 0.17'' for channel 2, 0.20'' for channel 3, and 0.35'' for channel 4. The output of the \texttt{Spec3} pipeline was calibrated spectral cubes which we used for the analysis in the following sections. 

\begin{figure*}[!htb]
    \centering
    \includegraphics[scale=0.7]{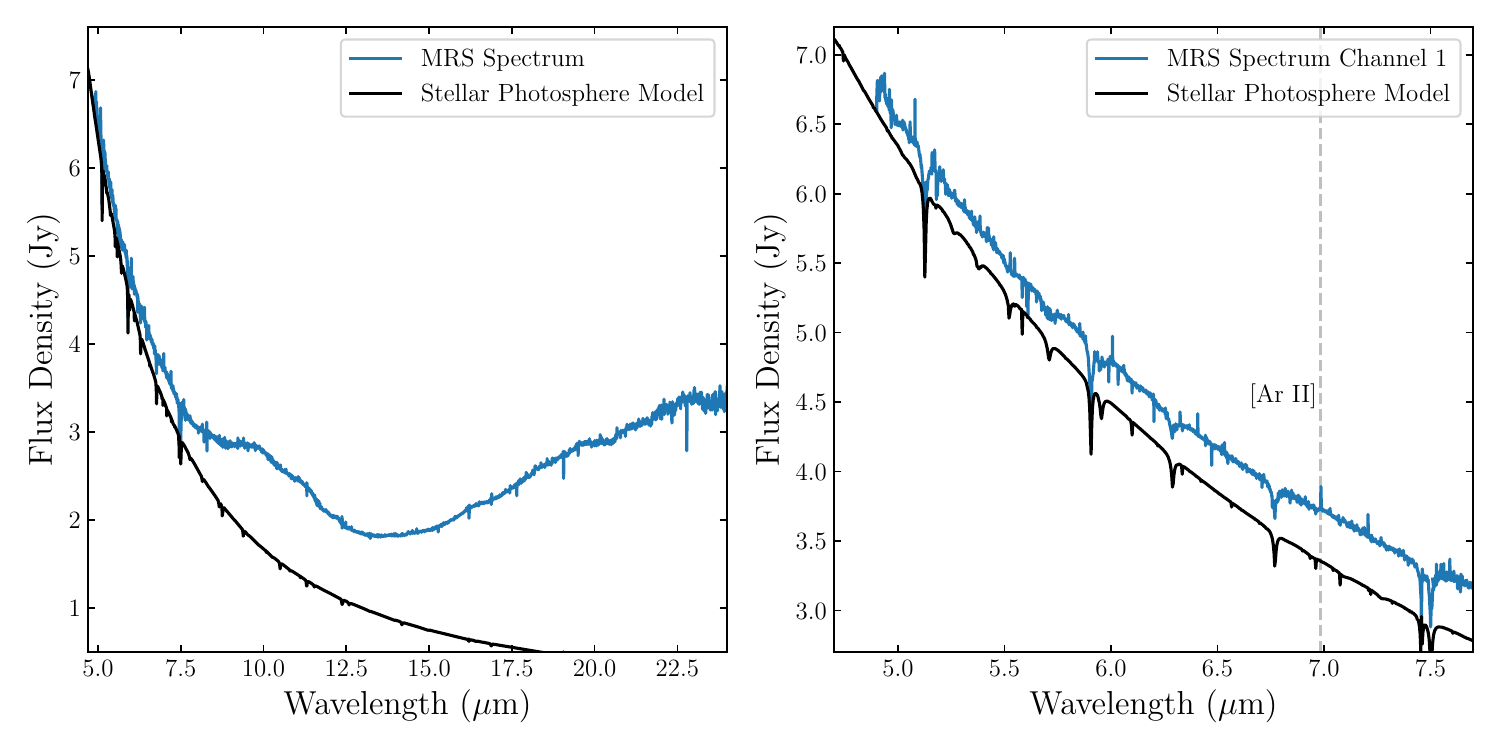}
    \caption{ Left: MRS spectrum of the $\beta$ Pic unresolved point source of channels 1-4. The feature from $\sim$8-12 $\text{\textmu}$m is from emission from silicate grains. Right: Channel 1 MRS spectrum of the $\beta$ Pic central point compared with a stellar photosphere model for the $\beta$ Pic star from \cite{Lu22} that was best fit to a near-IR spectrum. This shows an infrared excess from 5 to 7 $\text{\textmu}$m which can be explained by hot dust in the system. The vertical dashed grey line shows the detection of [Ar II] emission at 6.986 $\text{\textmu}$m. The other spikes in the spectrum do not appear to be real emission lines. This is because they only cover one wavelength bin, unlike the [Ar II] line which covers 5. Furthermore, these spikes also only appear after applying the RSRF and are well aligned to noise features seen in the spectrum of N Car. They then likely result from noise that is injected by performing the RSRF with N Car. The [Ar II] line, however, is present both before and after applying the RSRF.  }
    \label{fig:Spectrum}
\end{figure*}

\subsection{PSF Subtraction}\label{sec:PSF_sub}
To search for resolved disk structure as well as $\beta$ Pic b in our MRS data, we
used N Car to perform classical reference differential imaging (RDI) PSF 
subtraction on the calibrated spectral cubes. Because of position repeatability issues of the dichroic grating wheel assembly (DGA) of the MRS \citep{Patapis23}, the science star $\beta$ Pic and reference 
star N Car were not exactly centered in the same location on the detector. The position repeatability of the DGA wheel is $\sim$30 mas radially \citep{Patapis23}. 
 To achieve the best contrast performance from PSF subtraction, we aligned the PSF reference N Car with the PSF of $\beta$ Pic prior to subtracting. 

We first measured the position of the center of the PSF of $\beta$ Pic and N Car in the spectral cubes by fitting a 2D Gaussian to each wavelength slice and taking the median of the best fit center of all the wavelength slices. We did this for each sub-band separately (1A, 1B, etc.) because not all sub-bands were observed simultaneously, leading to position changes of the PSF in the spectral cube between the different sub-bands. We then calculated the difference between the position of the center of the PSF of $\beta$ Pic and N Car and shifted and interpolated the N Car PSF to the center location of the $\beta$ Pic PSF using the SciPy \texttt{ndimage.shift} function. After aligning the N Car PSF to the $\beta$ Pic PSF, we scaled the N Car PSF to have the same spectrum of the unresolved point source of $\beta$ Pic. We scaled to the spectrum of $\beta$ Pic rather than a photosphere model because there is an unresolved infrared excess in the central point source as shown in Figure \ref{fig:Spectrum} and scaling to just the photosphere model does not completely subtract the $\beta$ Pic PSF.

Once the N Car PSF was aligned and scaled to the flux level of $\beta$ Pic, we subtracted the N Car PSF from $\beta$ Pic at each wavelength slice of the spectral cube, producing a PSF subtracted spectral cube. We were left with large PSF subtraction residuals at the center location of the star, so to search for faint resolved disk structure as well as $\beta$ Pic b, we masked out spaxels within a 3 spaxel radius of the center of the PSF in channel 1 where the spaxel size is 0.13''. We also binned every 100 image slices in the data cube to increase the S/N of the individual image slices. 

\section{Results}
\subsection{ Spectrum of Central Point Source}\label{sec:unres_point}
To extract the spectrum of the unresolved point source of $\beta$ Pic, we performed aperture photometry at each wavelength slice of the spectral cubes. To find the center of the PSF, we collapsed each spectral cube along the wavelength axis and fit a 2-D Gaussian to the collapsed PSF. We placed the center of the circular aperture for each wavelength slice at the center of the PSF determined from the Gaussian fit. We then used a wavelength dependent aperture radius of 1.5$\times$ the FWHM of the PSF radius at each wavelength where we determined the FWHM of the PSF by taking a line cut through the center of the PSF in the direction orthogonal to the disk. We fit a Gaussian to the line cut of the PSF to measure the FWHM. At 5 $\text{\textmu}$m, the FWHM of the PSF from the Gaussian fit was 0.325''. This is larger than a diffraction limited PSF at 5 $\text{\textmu}$m, which would have a FWHM of 0.2''. \cite{Law23} similarly found that the size of the MRS PSF at 5 $\text{\textmu}$m is larger than what is expected from a diffraction limited PSF. After we extracted the flux within the circular aperture at each wavelength slice of the spectral cube, we applied an aperture correction to account for flux missed outside the extraction aperture. We applied an aperture correction for an extraction radius of 1.5$\times$ the PSF FWHM derived in \cite{Argyriou23} to the extracted spectra. We also applied a correction on the 1D spectrum, as described in \cite{Gasman23}, for the spectral leak at 12.2 $\text{\textmu}$m.

After performing aperture photometry on the spectral cubes, there were still noise sources, like fringing, present in the extracted spectra that were not completely removed from the pipeline processing. To refine calibration systematics, mitigate noise sources, and obtain the highest signal-to-noise spectrum of the unresolved point source of $\beta$ Pic, we used the observations of N Car to create a Relative Spectral Response Function (RSRF) and applied it to the spectrum of the unresolved point source of $\beta$ Pic. The RSRF was also used to align the different sub-bands. 

We extracted the spectrum of N Car using the exact same method as for $\beta$ Pic described above. We then fit the UBVGJHK$_s$ \citep{Cutri03,Reed03,Gaia18} photometry points for N Car with a $T=8800$ K, $log(g)=4.0, [Fe/H]=0$ BT-NextGen stellar photosphere model \citep{Allard11} using synthetic photometry and then divided this photosphere model by the N Car spectrum. This removed the stellar continuum and photospheric absorption lines from the N Car spectrum and provided an RSRF (consisting of detector effects present in the spectra that are present after pipeline processing). We then multiplied the RSRF into the $\beta$ Pic spectrum to remove the noise sources captured in the RSRF. In channel 1 ($\sim$5 to 7.5 $\text{\textmu}$m), applying the RSRF increased the S/N of the spectrum from $\sim$130 to $\sim$230. 

The final MRS spectrum of the unresolved point source of $\beta$ Pic is shown in Figure \ref{fig:Spectrum}. A surprising difference between the MIRI MRS spectrum shown here and the Spitzer IRS spectrum of $\beta$ Pic is that the sharp 18 $\text{\textmu}$m silicate feature seen with Spitzer \citep{Lu22}, is not present in our MRS spectrum. Here, we show the MRS spectrum and leave the analysis of the properties of the silicate feature and the discussion of the disappearance of the 18 $\text{\textmu}$m silicate feature for a future paper (Chen et al. \textit{in prep.}).

\subsection{5 $\text{\textmu}$m Infrared Excess}
We compared our extracted MRS spectrum to a model of the $\beta$ Pic photosphere from \cite{Lu22} to search for an IR excess at the shortest MRS wavelengths. This photosphere model is a BT-NextGen model fit to UBVRJHK$_s$ photometry as well as an IRTF SPEX spectrum of $\beta$ Pic from 0.7 to 2.5 $\text{\textmu}$m. This photosphere model was used throughout the rest of the analysis. 

At 5 $\text{\textmu}$m, we find an infrared excess that is 4$\%$ above the stellar photosphere model (see Figure \ref{fig:Spectrum}), potentially indicating the presence of dust within a stellocentric radius of $\sim$6.5 au (from the size of the extraction aperture) in the system. To determine the temperature of the dust emitting from 5 to 7.5 $\text{\textmu}$m, we subtracted the photosphere model from the MRS spectrum, smoothed it with a boxcar smoothing filter with a kernel of 3 data points, and then fit the channel 1 spectrum (4.9 to 7.5 $\text{\textmu}$m) with a blackbody. We also included the L-band photometric point of $\beta$ Pic from \cite{Bonnefoy13} of 3.454$\pm$0.003 mag. To subtract the photosphere from the L-band point, we calculated synthetic photometry using the stellar photosphere model of $\beta$ Pic and subtracted this value from the measured L-band photometry. Figure \ref{fig:blackbody_fit} shows the best fit blackbody to the 5 to 7.5 $\text{\textmu}$m excess. This blackbody fit gives a dust temperature of 500$\pm$20 K. There appears to be some residual structure in the spectrum shown in Figure \ref{fig:blackbody_fit}. This structure is likely due to an imperfect subtraction of the absorption lines in the stellar photosphere model. We think this because the structure in the spectrum (such as the dips at $\sim$5.6 and $\sim$7.5 $\text{\textmu}$m) are at the same wavelength as the stellar photosphere lines shown in Figure \ref{fig:Spectrum}. 

Even if the absolute flux calibration of the MRS spectrum was incorrect by 4$\%$, the shape of the spectrum still indicates an excess between 5.5-7.5 $\text{\textmu}$m. If we pin the shortest wavelength of the MRS spectrum (4.9 $\text{\textmu}$m) to the stellar photosphere model, there is still a $\sim3\%$ infrared excess at 5.5 $\text{\textmu}$m. Performing the same photosphere subtraction and blackbody fitting analysis, but with the spectrum pinned to the photosphere model at 4.9 $\text{\textmu}$m, gives a dust temperature of 370 K instead of 500 K. This still indicates the presence of dust in the inner few au of the system, just at a lower temperature. However, given that dust emission is spatially resolved at 5 $\text{\textmu}$m (see Section \ref{sec:Dust_spatial}), the excess we see in the spectrum at 5 $\text{\textmu}$m is likely real and not due to uncertainties in flux calibration. We are mainly interested in the temperature of the hot dust, so we only fit a blackbody to the 5 to 7.5 $\text{\textmu}$m excess and exclude longer wavelengths, including the silicate feature from $\sim$8 to 12 $\text{\textmu}$m. Fitting the MRS spectrum at wavelengths longer than 12 $\text{\textmu}$m would require a two component, two temperature model, and here we are only interested in the hot component. Adding a cold dust component does not affect the blackbody fit from 5 to 7.5 $\text{\textmu}$m. A complete modeling of the entire MRS spectrum including the silicate features will be done in an upcoming paper (Chen et al. \textit{in prep}). 

Assuming that the dust grains are blackbodies, in radiative equilibrium, and using the luminosity of $\beta$ Pic of 8.13 $L_{\odot}$ \citep{Bonnefoy13}, we calculated the blackbody location of the 500 K dust around $\beta$ Pic to be 0.9 au. This blackbody distance is a lower limit to the stellocentric distance of the grains, as smaller dust grains can be at higher temperatures further away from the star than blackbody grains. This is further discussed in section \ref{sec:Dust_spatial}. 

\begin{figure}[!htbp]
    \centering
    \includegraphics[scale=0.6]{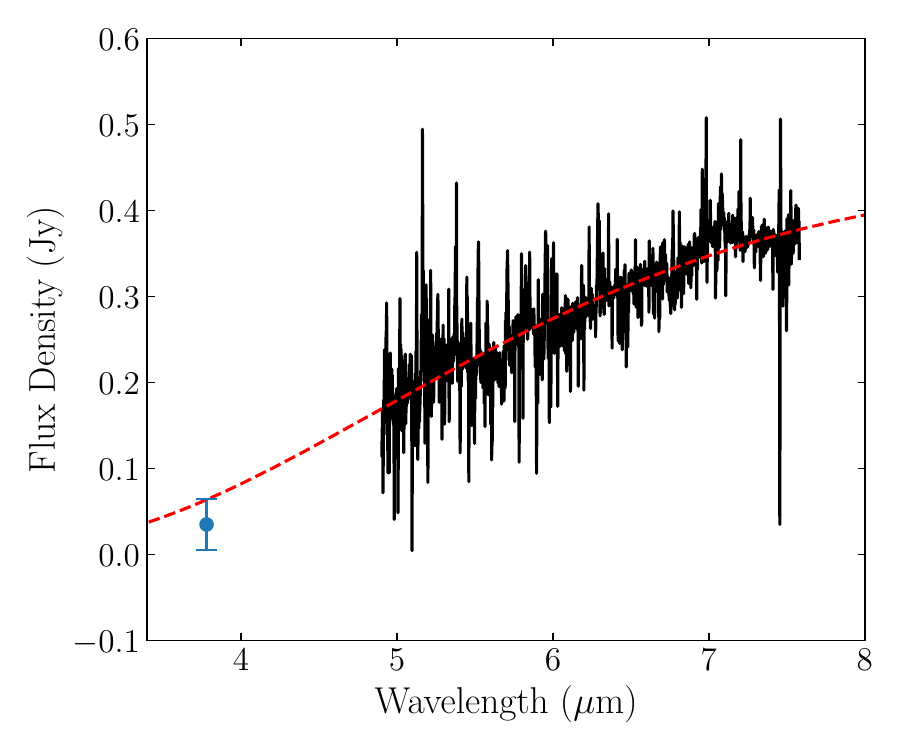}
    \caption{Best fit blackbody (red) to the photosphere subtracted MRS Channel 1 spectrum (black) and L-band photometric point (blue) from \cite{Bonnefoy13}. The best fit blackbody temperature is 500$\pm$20 K.}
    \label{fig:blackbody_fit}
\end{figure}

\subsection{Spatially Resolved Hot Dust}\label{sec:Dust_spatial}

\begin{figure*}[!htbp]
    \centering
    \includegraphics[scale=0.6]{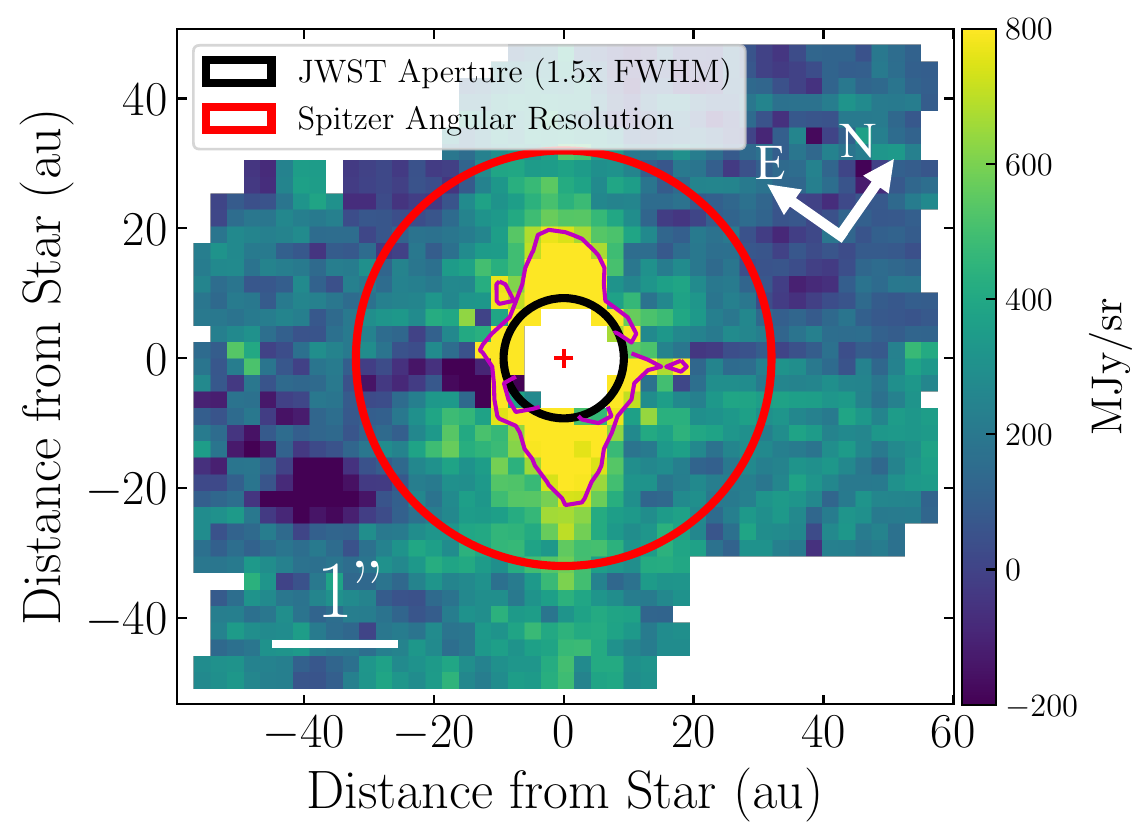}
    \caption{Binned 5.2 $\text{\textmu}$m slice of the PSF subtracted spectral cube showing the extended emission from hot dust. We apply an image normalization where v$_\text{min}=-200$ MJy/sr and v$_\text{max}=800$ MJy/sr to emphasize emission from a spatially extended disk. The purple line shows a 5$\sigma$ contour of the dust emission, the red circle shows the angular resolution of Spitzer at 5 $\text{\textmu}$m (assuming it is diffraction limited) and the black circle shows the extraction aperture used for the MRS. The red cross shows the location of the central star.}
    \label{fig:Extended_hot_disk}
\end{figure*}
With Spitzer, \cite{Lu22} found a tentative infrared excess at 5 $\text{\textmu}$m that could indicate the presence of hot dust within the inner parts of the system. Interferometric observations at H-band found evidence for an infrared excess that could either be from a population of hot dust (1500 K) at a distance of less than 4 au from the star, or scattered light from the outer part of the edge-on disk \citep{defrer12}. Here, we found additional evidence for the 5 to 7.5 $\text{\textmu}$m infrared excess seen by \cite{Lu22} in the spectrum of the unresolved point source (see Figure \ref{fig:Spectrum}). We confirmed that this population of hot dust emitting at 5 $\text{\textmu}$m is spatially extended in our PSF subtracted image cubes.

  The binned and PSF subtracted image slice at 5.2 $\text{\textmu}$m is shown in Figure \ref{fig:Extended_hot_disk} with the beam of Spitzer and the JWST extraction aperture overlaid. The purple line shows a 5$\sigma$ contour of the detected dust emission, indicating that we detect spatially resolved hot dust emitting at 5 $\text{\textmu}$m at a 5$\sigma$ level. Figure \ref{fig:Extended_hot_disk} illustrates that the excess thermal dust emission at 5 $\text{\textmu}$m was spatially unresolved given the angular resolution of Spitzer, but is now spatially resolved with JWST out to $\sim$ 20 au, where the disk flux density drops below the 5$\sigma$ level. By applying PSF subtraction to the MRS cubes, we have discovered a new population of spatially extended hot dust in the $\beta$ Pic debris disk emitting at 5 $\text{\textmu}$m.

 For dust at 10-20 au to emit at 5 $\text{\textmu}$m, the dust grains are likely to be sub-micron to micron-sized. In radiative equilibrium, the energy absorbed by a dust grain at a certain stellocentric distance is equal to the energy emitted from that dust grain, which can be written as 
 \begin{equation}
     \int_0^{\infty}Q_{abs}J_{\lambda}(T_*)d\lambda=\int_0^{\infty}Q_{abs}B_{\lambda}(T_d)d\lambda 
 \end{equation}
 where $Q_{abs}$ is the absorption coefficient at each wavelength for a specific material, $J_{\lambda}(T_*)$ is the mean intensity of the star, and $B_{\lambda}(T_d)$ is the Planck function representing the emission from the dust grain of a certain temperature $T_d$. 

 In the small grain limit, $Q_{abs}$ can be approximated as $Q_{abs}\propto (1/\lambda)$ where $\lambda$ is the wavelength of light. Using this small grain approximation and solving for the dust grain temperature as a function of stellocentric distance as in \cite{Jura98}, we get 
 \begin{equation}
     T_d=\left(\frac{R_*^2}{4D^2}\right)^{1/5}T_*
 \end{equation}
where  $T_d$ is the temperature of the dust, $R_*$ is the stellar radius, $D$ is the stellocentric distance of the dust, and $T_*$ is the effective temperature of the star. In the blackbody limit for larger dust grains, where $Q_{abs}=1$ at all wavelengths, the equation for dust temperature as a function of distance has the same form, however, the exponent changes from $1/5$ to $1/4$. We calculated the temperature as a function of stellocentric distance from $\beta$ Pic for dust grains in the blackbody and small grain limit assuming stellar parameters of $\beta$ Pic of $T_*=8000$ K \citep{Lu22} and $R_*=1.73$ $R_{\odot}$ \citep{Kervella03}. The temperature as a function of distance for dust grains around $\beta$ Pic in both the small grain and blackbody approximations is shown in Figure \ref{fig:Dust_temp}. 

To test whether the dust is better explained in the small-grain limit ($\sim$300-400 K, see Figure \ref{fig:Dust_temp}) versus the blackbody approximation ($\sim$100-160 K), we first summed the flux from all the pixels in the extended disk within the 5$\sigma$ contour shown in Figure \ref{fig:Extended_hot_disk}. This gives a total flux density from the extended emission at 5.2 $\text{\textmu}$m of 0.03$\pm$0.01 Jy. This is about 10 times less than the flux density of the spatially unresolved excess at the same wavelength shown in Figure \ref{fig:blackbody_fit}. We then estimated the number of dust particles that would be required to produce an observed flux density of 0.03 Jy at 5.2 $\text{\textmu}$m if the dust was in the blackbody approximation or the small-grain limit. At a stellocentric distance of 10 au, the dust temperature predicted by the blackbody approximation is 150 K. The observed flux from a single dust grain at a given temperature is given by the equation 
\begin{equation}
    F_{\lambda}=B(T,\lambda)\frac{a^2}{d^2}
\end{equation}\label{eqn:bb_flux}
where $B(T,\lambda)$ is the Planck function, $a$ is the radius of the dust grain, and $d$ is the distance of the dust to Earth. The number of dust particles required to produce the observed flux density from the extended emission at 5.2 $\text{\textmu}$m is then $N_{dust}=F_{obs}/F_{\lambda}$, where $F_{obs}=0.03$ Jy. For a dust temperature of 150 K in the blackbody limit and assuming a grain size equal to the wavelength of light of 5 $\text{\textmu}$m, we found $N_{dust}$=$1\times10^{34}$. We then used Equation 3 to calculate the predicted flux density at 20 $\text{\textmu}$m from this number of dust particles at 150 K. We found that the predicted 20 $\text{\textmu}$m flux is 450 Jy, which is much greater than, and thus inconsistent with, the observed flux at 20 $\text{\textmu}$m shown in Figure \ref{fig:Spectrum}. We performed the same calculation, but with dust in the small-grain limit ($a=1$ $\text{\textmu}$m) and at a temperature of 350 K (corresponding to a distance of 10 au as shown in Figure \ref{fig:Dust_temp}). We found that the number of dust particles required to produce the 0.03 Jy flux density was $N_{dust}=8\times10^{30}$ and that the predicted 20 $\text{\textmu}$m flux from this number of dust grains at a temperature of 350 K was 0.2 Jy. This predicted flux density is less than the observed excess at 20 $\text{\textmu}$m and thus consistent with the observations. Because the number of dust particles in the blackbody approximation required to produce the observed flux density from the spatially extended dust at 5 $\text{\textmu}$m is inconsistent with the observed flux density at longer wavelengths, the spatially extended 5 $\text{\textmu}$m excess is best explained by hotter sub-micron sized grains in the small-grain limit. This calculation yields the same conclusion regardless of the assumed grain size and dust temperature for all temperatures in the the small-grain and blackbody approximations within the red colored area in Figure \ref{fig:Dust_temp}.  
\begin{figure}[!htbp]
    \centering
    \includegraphics[scale=0.6]{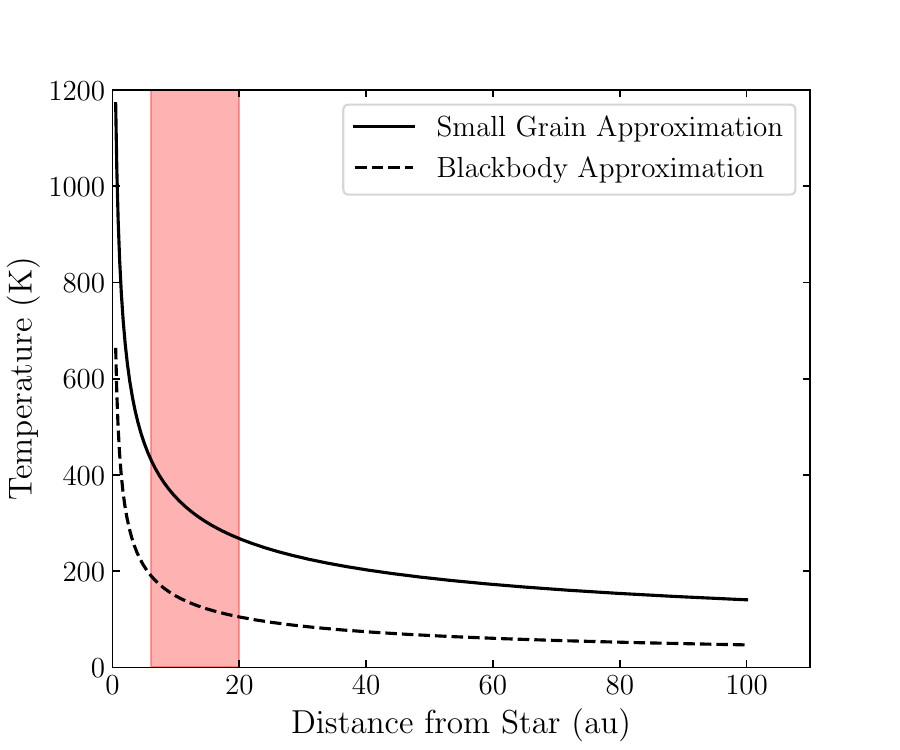}
    \caption{Dust temperature as a function of stellocentric distance from $\beta$ Pic for dust grains in the small grain approximation (black solid line) and for dust grains in the blackbody approximation (black dashed line). The red colored area shows the approximate location of the spatially resolved dust seen at 5 $\text{\textmu}$m (based on where the flux drops below a 5$\sigma$ detection).}
    \label{fig:Dust_temp}
\end{figure}

We thus cannot explain the spatially extended thermal emission at 5 $\text{\textmu}$m out to $\sim$20 au with blackbody grains, because at temperatures of 100-160 K, they do not emit significant flux at 5 $\text{\textmu}$m. The spatially extended dust seen at 5 $\text{\textmu}$m is better explained by dust grains in the small grain approximation. In the small grain limit ($2\pi a \ll \lambda$) at 5 $\text{\textmu}$m, the grains would have sub-micron radii and could be below the blowout size for the system (see Section \ref{sec:Dust_accret} for a calculation of the blowout size).


\subsection{The Detection of [Ar II] emission}
We searched for emission from atomic and molecular gas in our MRS spectrum of the unresolved point source. We did not have any clear detections of molecular gas emission (we search for \ce{H2O}, \ce{CO}, \ce{CO2}, and \ce{CH4}). We do, however, detect an emission line at 6.986 $\text{\textmu}$m that we believe to be due to [Ar II] (rest wavelength of 6.9853, \citealt{Ar_einsteinA}). To further verify this emission line, we reduced each dither position independently and found that the emission line is present in all four dither positions. We also did not detect the emission line in N Car's spectrum, suggesting that the emission line is not due to a instrument systematic or pipeline artifact. We fit the spatially unresolved component of the [Ar II] line with a Gaussian profile to determine its line flux, line width, and radial velocity. The argon emission line with the best fit Gaussian is shown in Figure \ref{fig:Argon}. 

From the best fit Gaussian, we measured the FWHM of the line to be 86$\pm7$ km/s and the line center to be at a velocity of 21.0$\pm$2.4 km/s. At the end of channel 1C (7.65 $\text{\textmu}$m), the spectral resolution of the MRS is $\sim$83 km/s \citep{Argyriou23}, so the [Ar II] line is consistent with being spectrally unresolved. The measured barycentric radial velocity of $\beta$ Pic is 20.0 $\pm$0.7 km/s \citep{gontcharov06}, which is consistent with the center velocity of the [Ar II] line. We calculated a line flux of 2.4$\times10^{-14}$ erg/s/cm$^{2}$.

\begin{figure}[!htbp]
    \centering
    \includegraphics[scale=0.45]{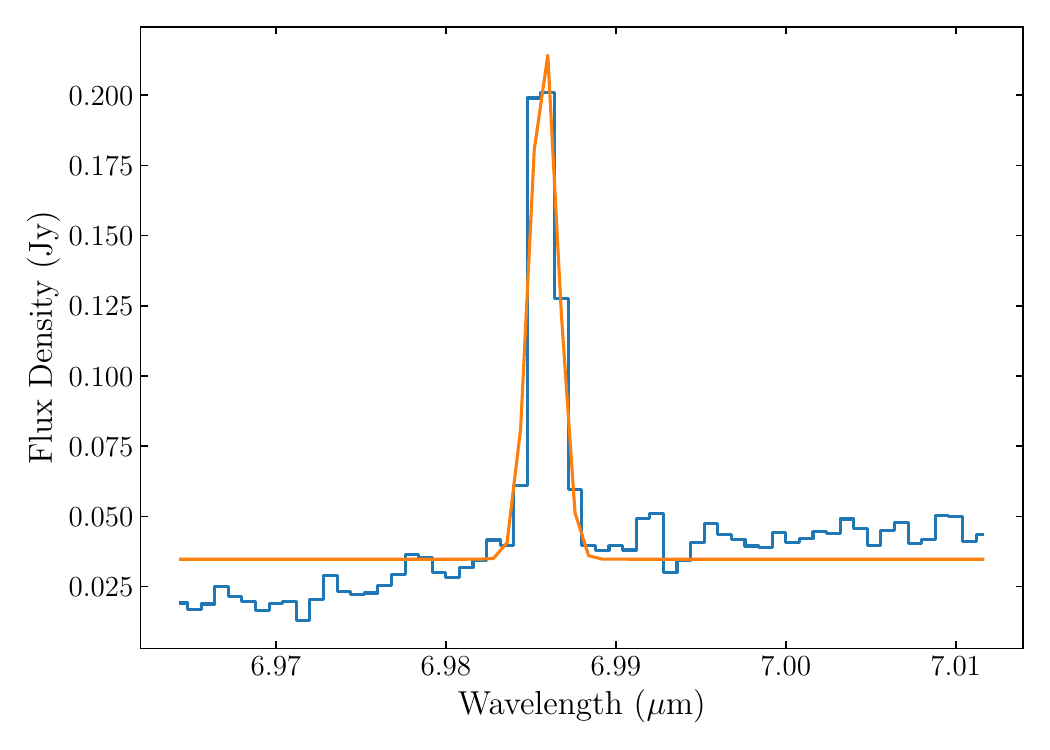}
    \caption{[Ar II] emission line at 6.986 $\text{\textmu}$m from the $\beta$ Pic unresolved point source with the best fit Gaussian overlaid. The best fit blackbody, as shown in Figure \ref{fig:blackbody_fit}, was subtracted from the spectrum in this figure. }
    \label{fig:Argon}
\end{figure}

We checked to see if the emission from [Ar II] is spatially resolved to determine if it is from circumstellar gas by subtracting a slice in the spectral cube outside of the argon line (6.9824 $\text{\textmu}$m) from the slice of the spectral cube at the peak of the [Ar II] line (6.9856 $\text{\textmu}$m). Doing this subtracted off the PSF of the unresolved point source as well as the continuum emission from the dust and planet ($\beta$ Pic b) so only emission from [Ar II] gas remains. The continuum subtracted cube slice of the [Ar II] emission is shown in Figure \ref{fig:argon_images}. In the 6.9856 $\text{\textmu}$m slice, there is a $\sim$5-10$\sigma$ detection of spatially resolved [Ar II] emission, where $\sigma$ is the standard deviation of the pixels in an annulus, at this wavelength slice, centered on the star with an inner radius of 1'' and an outer radius of 2.5''. By visual inspection, the argon appears to have a similar spatial distribution as the dust (see Figure \ref{fig:argon_images}), indicating that the argon is indeed a part of the $\beta$ Pic disk. To the best of our knowledge, this is the first detection of argon in the $\beta$ Pic disk as well as in any debris disk.

\begin{figure*}[!htbp]
    \centering
    \includegraphics[scale=0.7]{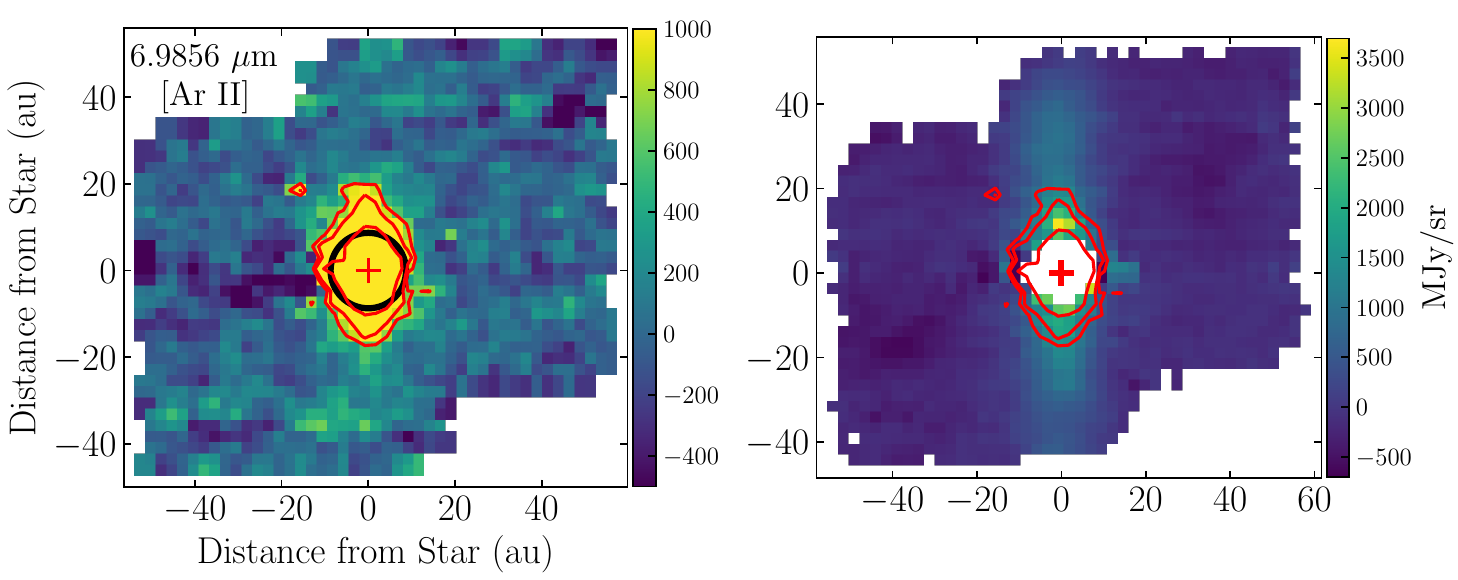}
    \caption{ Left: [Ar II] image created by subtracting the 6.9824 $\text{\textmu}$m slice (outside the argon line) from the 6.9856 $\text{\textmu}$m slice (the peak of the argon line). The red contours shown are 3, 5, and 10$\sigma$ contours. The black circle shows the FWHM of the JWST PSF at this wavelength, indicating that the [Ar II] emission is spatially resolved out to $\sim$20 au. The red cross shows the position of the central star. The North-East orientation of the images are the same as in Figure \ref{fig:Extended_hot_disk}. Right: binned PSF subtracted cube slice at 6.5 $\text{\textmu}$m with 3, 5, and 10$\sigma$ contours of the continuum subtracted [Ar II] emission overlaid. The point source on the NE side is $\beta$ Pic b.  }
    \label{fig:argon_images}
\end{figure*}

We used the line flux from the unresolved and resolved [Ar II] emission to estimate a total mass of [Ar II]. Under the assumption that the emission is optically thin, the total mass of [Ar II] is related to the line flux with the equation
\begin{equation}
    M_{\text{[Ar II]}}=\frac{4\pi d^2F m_s }{h \nu A_{ul}\chi_u}
\end{equation}

where $M_{\text{[Ar II]}}$ is the mass of [Ar II], $d$ is the distance to the star, $F$ is the line flux, $h$ is Planck's constant, $\nu$ is the frequency of the light, $A_{ul}$ is the Einstein A coefficient, $\chi_u$ is the fraction of atoms in the upper state, and $m_s$ is the mass of argon. We used an Einstein A coefficient of 5.3$\times10^{-2}$ s$^{-1}$ from \cite{Ar_einsteinA}. We calculated $\chi_u$ assuming the argon is in LTE and that it has an excitation temperature equal to the radiative equilibrium temperature profile shown for blackbody grains in Figure \ref{fig:Dust_temp}. This assumption of excitation temperature and LTE make the estimated [Ar II] mass highly uncertain because, with only one [Ar II] line, we have no knowledge of the excitation temperature of the gas or if it is in LTE. 

Since we expect the gas temperature to change as a function of disk radius, we calculated the line flux separately for the spatially resolved [Ar II] component. We did this by using rectangular extraction apertures of 2 pixels in height and the width was set by the 3$\sigma$ contour of the [Ar II] emission shown in Figure \ref{fig:argon_images}. We placed these apertures centered at stellocentric distances of 10 and 15 au on both the northeast and southwest sides of the disk. We then fit the spectra from each aperture with a Gaussian and integrated the best fit Gaussian to obtain a line flux as described above. We then summed together the line fluxes from the two sides of the disk at the same stellocentric distance, giving a total [Ar II] line flux at 10 and 15 au. We then used Equation 4 and excitation temperatures at 10 and 15 au equal to the radiative equilibrium temperatures shown in Figure \ref{fig:Dust_temp} (150 and 125 K respectively) to calculate the [Ar II] mass at each stellocentric distance. 

For the unresolved [Ar II] emission, the location and LTE temperature of the gas is unknown because the spectral resolution of the MRS is not high enough to resolve gas kinematics. We then assumed that the unresolved [Ar II] has a temperature (180 K) corresponding to a stellocentric distance equal to the PSF FWHM at 6.986 $\text{\textmu}$m, which is $\sim$7 au. This assumption is likely incorrect, however, we made it because it then gives an upper limit to the total gas mass in the unresolved point source. We then computed the total mass by summing the calculated mass at each stellocentric distance. This gives a total [Ar II] mass of 1$\times 10^{-3}$ M$_{\oplus}$. Given the uncertainty on the inner-edge of the argon disk and the excitation of the gas, there is likely at least one to three orders of magnitude of uncertainty on this [Ar II] mass.

\subsection{$\beta$ Pic b}
We also searched for $\beta$ Pic b in our binned PSF subtracted spectral cubes as it was predicted to be at an angular separation of 0.54'' on 2023, January 11, the day of our MRS observations \citep{Lacour21, Wang21}. We detect a point source in channels 1A, 1B, and 1C in the PSF subtracted cubes. Slices from the PSF subtracted cubes from all three sub-bands of channel 1, showing the detection of a point source are shown in Figure \ref{fig:Planet_gal}. We detect this point source throughout channels 1A, 1B, and about half of channel 1C (up to $\sim$6.8 $\text{\textmu}$m). At longer wavelengths, thermal emission from the edge-on disk becomes too bright and we cannot recover the point source from the emission of the co-spatial disk (see Figure \ref{fig:10_mic_disk}). At the three different wavelength slices shown in Figure \ref{fig:Planet_gal}, we fit a 2D Gaussian to the point source and measured its separation and position angle from the PSF center of the star at each sub-band. We computed the median of the position angle and angular separation of the point source for the three image slices from the three different sub-bands and took their standard deviations to be the uncertainty. We measured a separation of the point source of 0.55$\pm$0.02'' and a position angle of 31$\pm$1 degrees. The predicted location of $\beta$ Pic b on the date of our observations, based on high-precision GRAVITY astrometry measurements, was a separation of 0.540$\pm$.003'' and a position angle of 31.571$\pm0.006$ degrees \citep{Lacour21,Wang21}, which is consistent with the location of the detected point source in our PSF subtracted spectral cubes. 

\begin{figure*}[!htbp]

    \centering
    
    \includegraphics[scale=0.55]{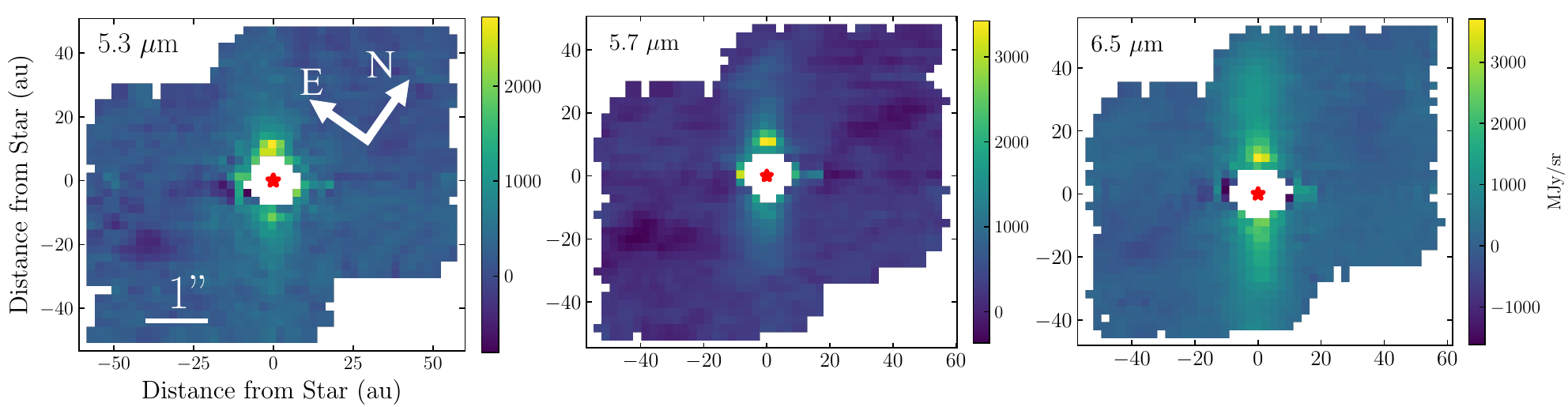}
    \caption{Binned PSF subtracted image slices showing the detection of a point source at the predicted location of $\beta$ Pic b (separation=0.54'', PA=31.57$^{\circ}$) (\citealt{Wang21}, \citealt{Lacour21}). The left shows a binned slice from channel 1A, the middle 1B, and the right 1C. The red star indicates the position of the central star determined from the 2D Gaussian fit to the PSF. The pixels within a 3 pixel radius of that center position are masked for clarity. The S/N of the brightest pixel in the point source, going from left to right, is 5.1, 6.3, and 5.5}
    \label{fig:Planet_gal}
\end{figure*}

We computed contrast curves by masking out the planet and then calculating the standard deviation (1$\sigma$) in concentric annuli of a 2 pixel width in the binned PSF subtracted image slices as a function of angular separation from the center of the PSF. We then divided these standard deviations as a function of angular separation by the spectrum of the unresolved point source to estimate the contrast performance. We did this both in the direction of the disk, to determine how well we recover $\beta$ Pic b, and also outside of the disk in order to show the general contrast performance of our PSF subtraction with the MRS. To compute the contrast curves in the direction of the disk, we included all columns of pixels that included disk emission at the 3$\sigma$ level. The contrast curves in the direction of the disk and outside of the disk for the three wavelength slices in Figure \ref{fig:Planet_gal} are shown in Figure \ref{fig:contrast_curves}. The contrast curves computed in the direction of the disk show that we are achieving a 5$\sigma$ contrast of $\sim1\times10^{-4}$ at a separation of 1.0 arcseconds at 5.3 and 5.7 $\text{\textmu}$m. At 6.5 $\text{\textmu}$m, we achieve a 5-$\sigma$ contrast of $3\times10^{-4}$. The decrease in the contrast performance from 5.7 and 6.5 $\text{\textmu}$m between separations of 0.6-2.5 arcseconds seen in the left plot of Figure \ref{fig:contrast_curves} is likely because the thermal emission from the disk is brighter and extended out to larger stellocentric distances at longer wavelengths. We also computed contrast curves with the extended disk emission masked out. This was done to estimate the general contrast performance of our PSF subtraction technique with the MRS without contamination from the extended disk emission. These contrast curves are shown in the right panel of Figure \ref{fig:contrast_curves}. With the disk emission masked, our PSF subtraction and binning method achieves 5$\sigma$ contrast levels of $\sim1\times10^{-4}$ at all three wavelengths at separations $\ge$ 1 arcsecond.  

We estimated the contrast of $\beta$ Pic b by dividing its spectrum, extracted from aperture photometry in Section \ref{sec:bpicb}, by that of the central unresolved point source (star+disk) from Section \ref{sec:unres_point}. We then compared this contrast to the calculated contrast curves shown in Figure \ref{fig:contrast_curves} to estimate the S/N of $\beta$ Pic b. The S/N of $\beta$ Pic b at 5.3, 5.7, and 6.5 $\text{\textmu}$m in our binned MRS image cubes is 5.1, 6.3, and 5.5 respectively.

\begin{figure}[!htbp]
    \centering
    \includegraphics[scale=0.5]{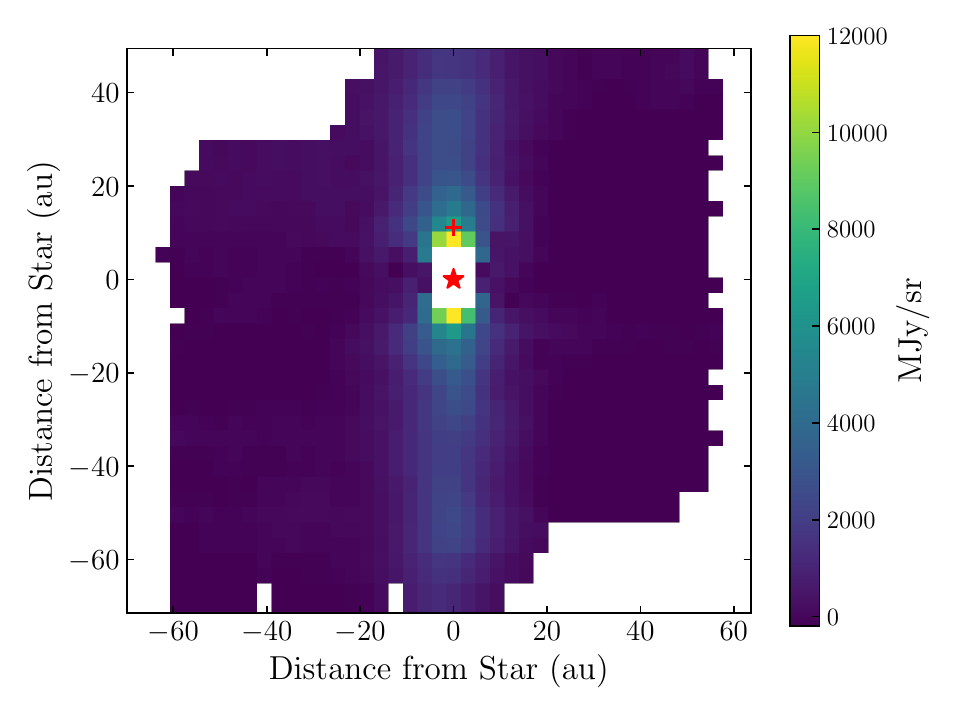}
    \caption{PSF subtracted and binned slice of the channel 2B data cube at 10 $\text{\textmu}$m. The red star indicates the position of the central star and the red cross indicates the position of $\beta$ Pic b at the position angle and separation we found from the channel 1 data cubes. The pixels within a 2 pixel radius of the PSF center have been masked for clarity. The image slice is dominated by thermal emission from the disk and we do not detect a clear point source on one side of the disk as we did in channel 1. The orientation of this image relative to North is the same as those shown in Figure \ref{fig:Planet_gal}.}
    \label{fig:10_mic_disk}
\end{figure}

\begin{figure*}[!htbp]
    \centering
    \includegraphics[scale=0.65]{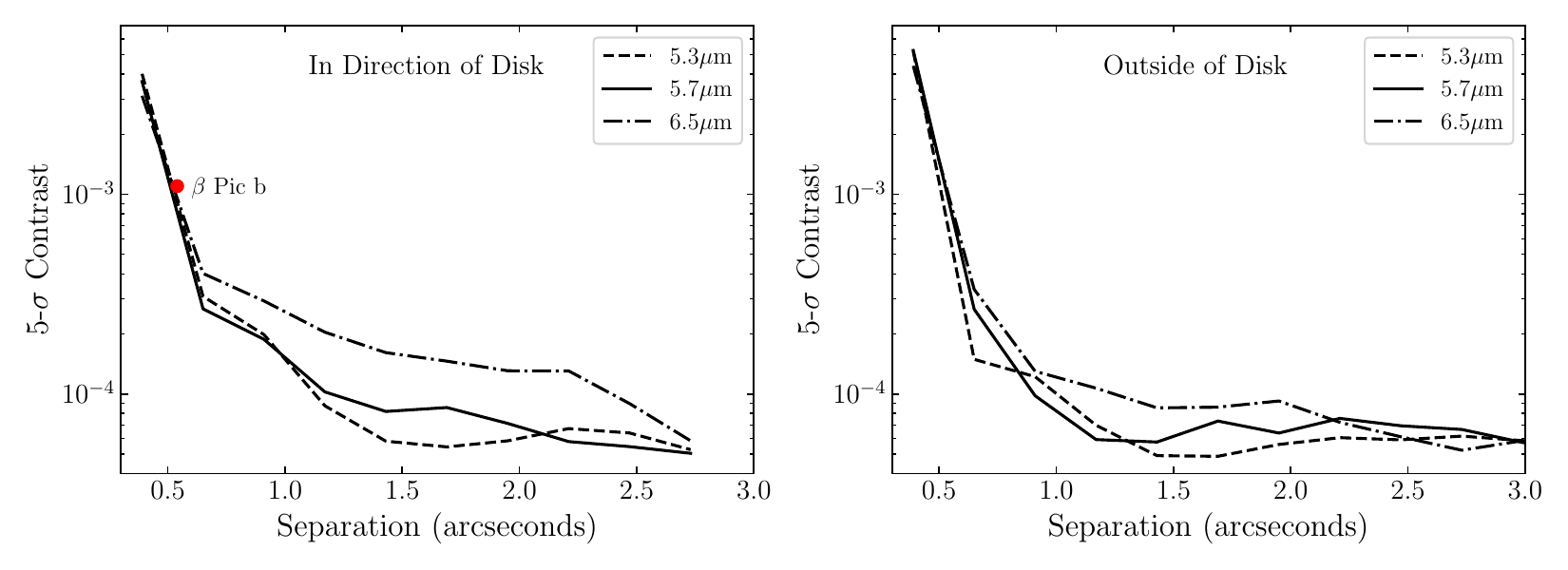}
    \caption{Left: contrast curves in the direction of the disk from the binned PSF subtracted image cubes at wavelength slices in channels 1A, 1B, and 1C (same slices shown in Figure \ref{fig:Planet_gal}). The red point shows the contrast of $\beta$ Pic b calculated at 5.7 $\text{\textmu}$m from our data. We do not show the contrast of $\beta$ Pic b at the other two wavelengths for clarity. Right: contrast curves from the binned PSF subtracted image cubes at the same wavelength slices as in the left plot but over the entire field of view with the extended emission from the disk masked. This shows the contrast performance of our PSF subtraction with the MRS without contamination from the extended disk emission.}
    \label{fig:contrast_curves}
\end{figure*}

\subsection{Cross-correlation}
We also searched for $\beta$ Pic b using cross-correlation methods with a template planetary atmosphere spectrum. This method involves cross correlating the spectrum of each spaxel of the data cube with a planetary atmosphere model with various molecular absorption lines to search for molecules present in the atmosphere of the planet. This creates a cross-correlation map, where each spaxel in the IFU cube has one computed cross-correlation value. At the location of the planet in the spectral cube, where there is molecular absorption lines from its atmosphere, the cross-correlation function will peak, while the spaxels in the cube that do not contain a planet with the absorption features of the specific molecule being tested will appear as a featureless background.

For the atmosphere template spectrum, we generated \texttt{petitRADTRANS} \citep{molliere19} models at their full spectral resolution with absorption from various molecules. We used the physical parameters of $\beta$ Pic b determined by the \cite{Nowak20} ($T_{eff}=1742$ K, $log(g)=4.34$). For each model we generated, we only included absorption from one molecular species to search specifically for each individual species in the atmosphere of $\beta$ Pic b. We searched for \ce{H2O}, \ce{CO}, \ce{CH4}, \ce{CO2}, and \ce{NH3} in all wavelength channels. For each molecule, we input a mass fraction of $1\times10^{-4}$ into the atmosphere model. We tried a range of mass fractions from $1\times10^{-3}$ to $1\times10^{-6}$ and found that the result of the cross-correlation detections did not change for any molecule.

We first removed the continuum from the atmosphere model by fitting the model with a B-spline of 10 breakpoints and subtracting off this continuum fit, leaving only the molecular absorption lines. The same was done for the spectrum of each spaxel in the spectral cube, to remove the continuum spectrum. Once we removed the continuum of the atmosphere model and each spaxel, we then cross-correlated the atmosphere model templates with the continuum-subtracted spectra of each spaxel to calculate a cross-correlation value. We then estimated the S/N per spaxel of any potential cross-correlation detection by dividing by the standard deviation of cross-correlation values of all the spaxels, which is  similar to what is done by \cite{Malin23}.

This produces a S/N detection map where each spaxel in the image cube has a S/N value. Figure \ref{fig:cross_Cor} shows the cross-correlation map of the full Channel 1 cube with a \texttt{petitRADTRANS} atmosphere model containing \ce{H2O} as the only absorbing molecular species. We find a bright pixel in the cross-correlation map at the same location as the brightest pixel of the point source recovered in the PSF subtracted images slices. Out of all the molecules and MRS channels we tested, \ce{H2O} in channel 1 was the only one that resulted in at least one pixel above a S/N value of 5.
The bright pixel in our detection map has a S/N value of 5.7. \cite{Malin23} simulated cross-correlation techniques using simulated MRS observations of the $\beta$ Pic system and similarly found that \ce{H2O} was the only molecule that yielded a significant S/N value for $\beta$ Pic b \citep[See Figure C.2. of][]{Malin23}.

Previous studies that use cross-correlation techniques with near-infrared ground-based data have detections of planets that cover multiple IFU spaxels \citep{Hoeijmakers18}, while here we only detect one spaxel with a significant cross-correlation value, which lines up with the position of the planet from the PSF subtracted image slices. Doing the same cross-correlation method with all the molecular species listed above did not result in any spaxel in the cubes having a cross-correlation value greater than 5 in any of the MRS channels (1-4), so finding a spaxel with a S/N value of 5.7 using \ce{H2O} does not appear to be random chance. Furthermore, with 1498 spaxels in the channel 1 data cube, it is unlikely that the cross-correlation method would have randomly selected the same spaxel as the point source in the PSF subtracted cubes to have a significant cross-correlation value.

The detection of a point source through PSF subtraction at the predicted location of $\beta$ Pic b based on orbit fits from \cite{Lacour21} in three MRS sub-bands as well as the tentative detection of water through cross-correlation at the same location as the point source provides enough confidence for us to conclude that we are in fact detecting $\beta$ Pic b in the MRS data. 

\begin{figure}[!htbp]
    \centering
    \includegraphics[scale=0.55]{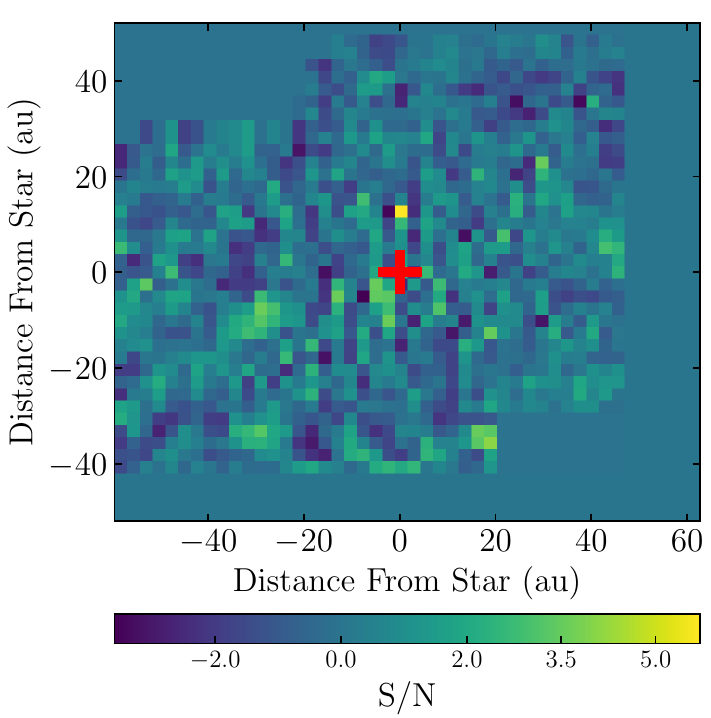}
    \caption{Cross-correlation map of the Channel 1 IFU cube with a atmosphere template spectrum with water being the only molecular absorber. We detect a bright pixel with a cross-correlation S/N value of 5.7 at the same pixel location as the brightest pixel of the point source shown in Figure \ref{fig:Planet_gal}. The red cross shows the position of the central star. }
    \label{fig:cross_Cor}
\end{figure}

\subsection{Spectrum of $\beta$ Pic b}\label{sec:bpicb}
We performed aperture photometry on $\beta$ Pic b at each slice of the binned PSF subtracted spectral cubes  to extract a spectrum of the planet. We found that the co-located spatially extended dust emission contaminates the spectrum of the planet. In order to figure out the best strategy for subtracting the contribution of the dust, we performed an injection recovery test using the N Car PSF. We first scaled the PSF of N Car in each wavelength slice such that the brightest pixel of the N Car PSF matches the brightest pixel of $\beta$ Pic b at the first wavelength slice in each sub-band. In the last wavelength slice of each sub-band, the brightest pixel in the injected point source was within 5$\%$ of that of $\beta$ Pic b. The spectrum of N Car was not changed aside from the scaling. We then injected the scaled N Car PSF on the opposite side of the disk at the same separation as $\beta$ Pic b before performing PSF subtraction. We PSF subtracted each spectral cube with N Car injected the same way as described in Section \ref{sec:PSF_sub} and produced a binned and PSF subtracted spectral cube, which is shown in Figure \ref{Injected_planet_images}.

\begin{figure*}[!htbp]
    \centering
    \includegraphics[scale=0.5]{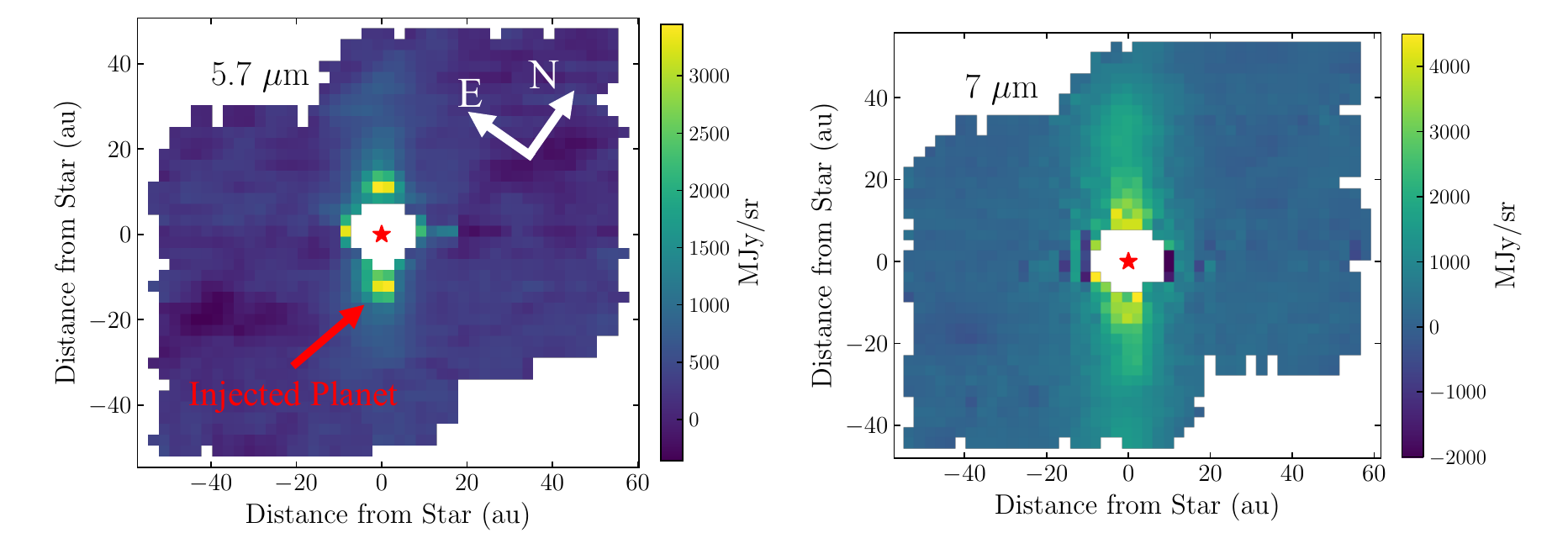}
    \caption{Left: same as the 5.7 $\text{\textmu}$m slice shown in Figure \ref{fig:Planet_gal} but now with N Car injected as a fake planet on the opposite side of the disk. The point source on the northeast side of the disk is $\beta$ Pic b. Right: same as left but at the 7 $\text{\textmu}$m slice, where there is no clear detection of the injected point source or of $\beta$ Pic b.}
    \label{Injected_planet_images}
\end{figure*}

We performed aperture photometry on the injected planet and subtracted off the disk background by placing a background aperture of the exact same size at a further separation of 0.8'' on the same side of the disk as the injected planet. This location of background aperture was chosen because it is the closest location to the injected planet without the extraction and background apertures overlapping. We used an aperture radius of 0.5$\times$ FWHM of the PSF because we found that using a larger aperture includes more thermal emission from the disk, which results in a worse recovery of the injected spectrum. We created our own aperture correction for this aperture size using N Car. We binned the spectral cube of N Car the same way as the PSF subtracted cubes, and then extracted the spectrum of N Car the same way as described above for the unresolved point source of $\beta$ Pic. We also extracted the binned spectrum of N Car using an aperture of 0.5$\times$ FWHM and divided the two to derive an aperture correction for an aperture size of 0.5$\times$ FWHM. Using this background subtraction method and extraction aperture size, we were able to recover the general shape of the spectrum of the injected point source, however, we were unable to recover the absolute flux density values. The injected and recovered spectrum for channel 1B is shown in Figure \ref{fig:Inject}.  At the same wavelength that we were unable to detect $\beta$ Pic b ($\sim$7 $\text{\textmu}$m), we were unable to recover the injected point source as well (see Figure \ref{Injected_planet_images}). This is potentially because beyond this wavelength, the thermal emission from the disk is bright enough that it dominates over the planet flux, and we cannot recover the planet at longer wavelengths. We calculated the difference between the injected and recovered spectrum and then subtracted this off of the extracted spectrum of $\beta$ Pic b. This was to remove the excess flux from the co-spatial disk that was not subtracted with the background aperture, as indicated by the injection recovery test.

\begin{figure}[!htbp]
    \centering
    \includegraphics[scale=0.45]{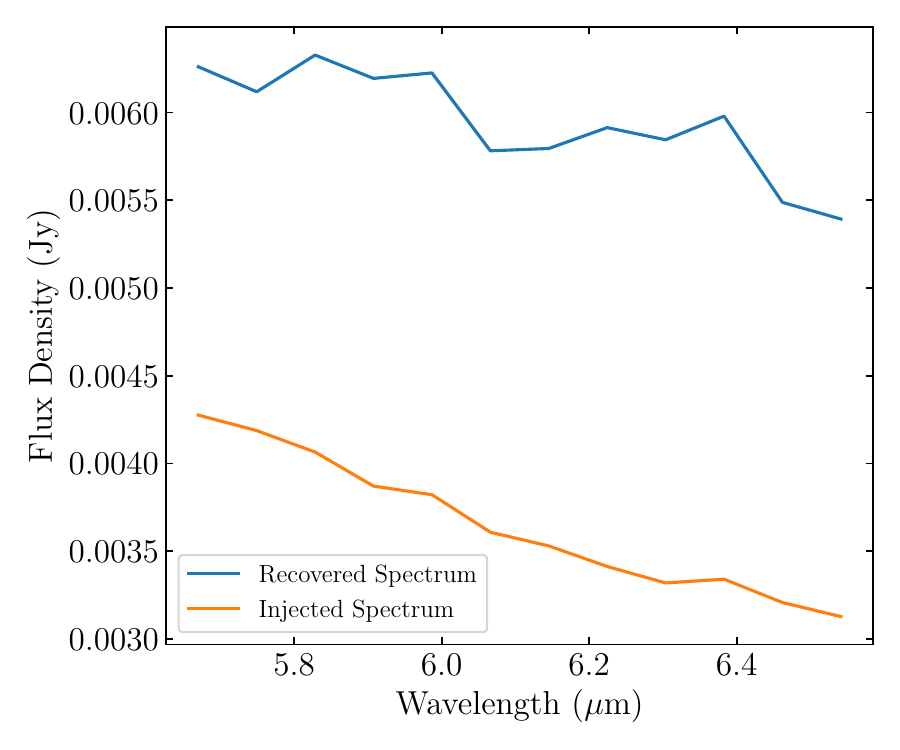}
    \caption{Channel 1B spectrum of the injected point source (orange) and the spectrum we recover from the injected point source (blue). We generally recover the overall slope of the injected spectrum, however, we are unable to recover the absolute flux density of the spectrum. Even with subtracting off the flux density in a background aperture in an attempt to remove contamination from the spatially extended disk, the recovered spectrum still has greater flux density than the injected spectrum by a factor of $\sim$1.3-1.4.}
    \label{fig:Inject}
\end{figure}

 We used the same background aperture, same aperture size, and same aperture corrections for extracting the spectrum of $\beta$ Pic b as we did for the injected fake planet. We placed the extraction aperture on the center of $\beta$ Pic b measured by the 2D Gaussian fit for each sub-band. We estimated the uncertainty on the spectrum of $\beta$ Pic b from the contrast curves shown in Figure \ref{fig:contrast_curves}. That is, the standard deviation of the annulus containing containing the separation of $\beta$ Pic b at each wavelength slice is used as the uncertainty. The final spectrum of $\beta$ Pic b, along with an ExoREM \citep{Charnay18,blain21} planetary atmosphere model, is shown in Figure \ref{fig:b_pic_b_spec}. 

  \begin{figure}[!htbp]

    \centering
    \includegraphics[scale=0.5]{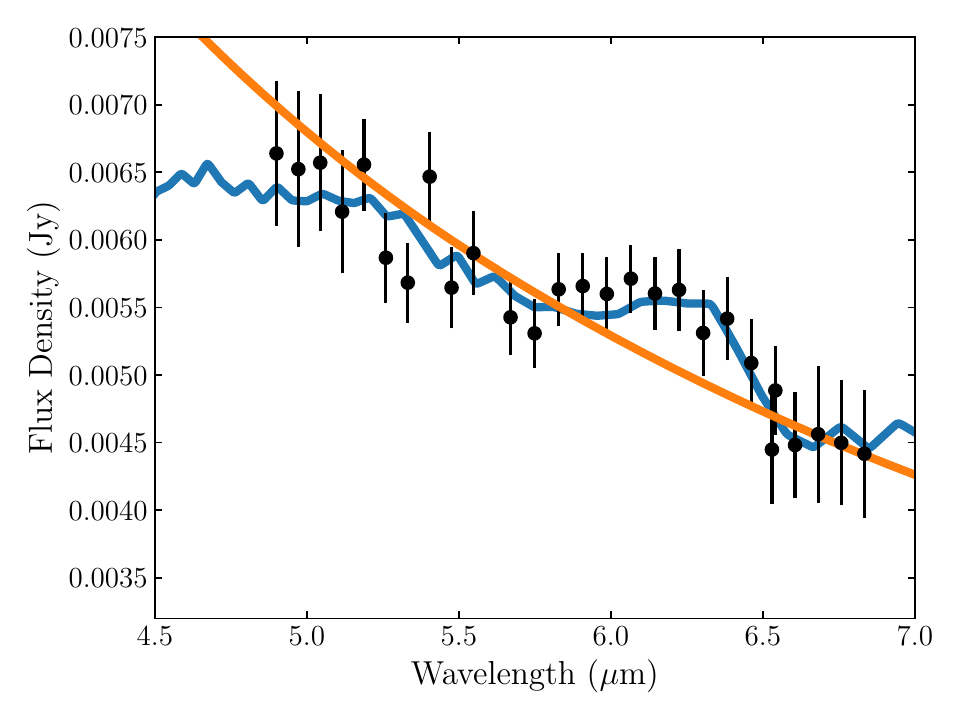}
    \caption{MIRI MRS spectrum of $\beta$ Pic b (black) with best fit ExoREM model overplotted (blue) and a best fit blackbody (orange).The broad absorption bands from 5.1 to 6.0 microns and from 6.4 to 6.8 microns are due to absorption from water vapor. The local maximum in the spectrum at 6.3 microns is due to a relative absence of water vapor lines near this wavelength. The spectrum is better fit by the atmosphere model that contains water absorption rather than the blackbody with no molecular features since the blackbody cannot reproduce the shape of the spectrum between 6 and 6.5 $\text{\textmu}$m. The spectrum showing signs of water absorption is consistent with the tentative detection of water from the cross-correlation technique.}
    \label{fig:b_pic_b_spec}
\end{figure}

 \subsection{Atmospheric Model Fitting}

 We used the \texttt{Species} package \citep{Stolker20} along with three different sets of model grids of synthetic spectra to fit the spectra of $\beta$ Pic b. The first was the ATMO model grid \citep{Tremblin15} presented in \cite{Petrus23}. The second model grid we used was the DRIFT-PHOENIX model grid \citep{Helling08} and the third was the ExoREM grid \citep{Charnay18,blain21}.
 The model parameters and the prior ranges for all three model grids are shown in Table \ref{tab:Bpic_b Parameters}. We used uniform priors for each parameter in each of the model grids. In the modeling fitting, we included our MRS spectrum, the GRAVITY K-band spectrum from \cite{Nowak20}, and the GPI YJH spectra from \cite{Chilcote17}. We also included photometric measurements from Magellan and Gemini NICI \citep{Males14} and VLT/NACO \citep{Bonnefoy13}. Since we were unable to recover the absolute flux density of the injected planet in the injected recovery test because of the co-spatial thermal dust emission, we left the overall flux scaling of the MRS spectrum as a free parameter in the model fitting, similar to what was done in \cite{Kammerer21} to align spectra from GPI and GRAVITY for the substellar object HD 206893 B. We weighted each spectrum and photometric point in the log-likelihood function in the \texttt{Species} model fitting such that each data set, spectra and photometric, had the same weighting. This was to ensure that the model fitting is not dominated by the GRAVITY spectrum which contains the most data points. We then infered the posterior distribution of the parameters using nested sampling with \texttt{PyMultiNest} \citep{buchner14,Feroz09,Feroz19}. The MRS spectrum of $\beta$ Pic b is shown in Figure \ref{fig:b_pic_b_spec} with the best fit ExoREM model. The posterior distributions for the model fits are shown in the Appendix.

Figure \ref{fig:species_fits} shows the best fit Drift-Phoenix, ATMO, and ExoREM models to all of the spectra and photometry points. Table \ref{tab:Bpic_b Parameters} shows the best fit model parameters for each of the model grids. Both the ATMO and Drift-Phoneix model grids give consistent measurements within the uncertainties for $log(g)$ and effective temperature. The ExoREM grid gives a lower effective temperature by $\sim$200 K than the other two model grids. The temperatures from the ATMO and Drift-Phoenix grids are within the uncertianties of each other. Fits with ExoREM models have given lower effective temperatures than other model grids in other fits in the literature as well (e.g., \citealt{Nowak20,Kammerer21}). All three model grids produce inconsistent metallicities. The ATMO model grid gives a best fit metallicity that is sub-solar, while the Drift-Phoenix model grid gives a super-solar metallicity and the ExoREM grid gives a larger super-solar metallicity (see Table 3). The uncertainty in the metallicity from the ATMO grid does include solar metallicity, although the uncertainties of the metallicity between the three model grids do not overlap, suggesting that we do not constrain the metallicity of the planet well with this method of grid model fitting.

To test how the new MRS data affects the constraints on the atmosphere of $\beta$ Pic b from our modeling, we repeated the grid model fitting excluding the MRS spectrum. We find that the addition of the MRS spectrum does not significantly change the results from the grid model fitting. All of the atmospheric parameters from fitting each of the three model grids are consistent with each other (within the uncertainties) when we performed the model fits with and without the MRS spectrum. The C/O ratio from ExoREM and ATMO model fits is similarly not significantly changed. Excluding the MRS spectrum with the ExoREM grid yielded a C/O ratio of 0.38$^{+0.10}_{-0.02}$, which is consistent with the C/O ratio obtained when including the MRS spectrum (See Table \ref{tab:Bpic_b Parameters}). Excluding the MRS spectrum from  he ATMO model grid fit gives a C/O ratio of 0.38$^{+0.10}_{-0.06}$, which is consistent with the ATMO fit that included the MRS spectrum. The shape of the MRS spectrum of $\beta$ Pic b is therefore consistent with what is predicted by the atmosphere model fits to the near-IR spectra and photometry.

\begin{table*}[!htbp]
    \centering
    \caption{Best fit Model Parameters for $\beta$ Pic b}
    \hspace{-1.2cm}
    \begin{tabular}{c c c c c c c }
    \hline
        Model Parameter& Drift-Phoenix& Drift-Phoenix Priors& ATMO& ATMO Priors & ExoREM&ExoREM Priors  \\
        \hline
         $T_{eff}$ (K)&1738$^{+16}_{-21}$& (1500,1800)& 1703$^{+37}_{-44}$&(1500,1800) & 1471$^{+20}_{-18}$&(1500,1800) \\
        $log(g)$ &4.03$^{+0.07}_{-0.07}$&(3,5)&3.98$^{+0.08}_{-0.12}$&(3,5)&3.71$^{+0.07}_{-0.08}$ &(3,5) \\
        $[Fe/H]$ &0.18$^{+0.07}_{-0.07}$ &(-0.6,0.3) &-0.12$^{+0.20}_{-0.19}$& (-0.6,0.6)&0.90$^{+0.07}_{-0.13}$&(-0.5,1.0)\\
        $C/O$& N/A& N/A&0.39$^{+0.10}_{-0.06}$&(0.3,0.7)&0.36$^{+0.13}_{-0.05}$&(0.1,0.8)\\
        Radius ($R_J$)&1.40$^{+0.04}_{-0.03}$&(0.5,2)& 1.39$^{+0.06}_{-0.06}$&(0.5,2)& 1.97$^{+0.05}_{-0.05}$&(0.5,2)\\
        $log(L/L_{\odot})$& -3.77$^{+0.01}_{-0.01}$&N/A& -3.81$^{+0.01}_{-0.01}$&N/A&-3.76$^{+0.01}_{-0.01}$&N/A
        \\
        $\chi^2_{red}$&5.5&N/A&5.9&N/A&3.3&N/A\\
        \hline
    \end{tabular}
    \begin{minipage}{14 cm}
    
      \vspace{0.1cm}
      \textbf{Note.} The $\chi^2_{red}$  value reported here is for all the photometry and spectra shown in Figure \ref{fig:species_fits}, not just the MRS spectra. 
     \end{minipage}
    \label{tab:Bpic_b Parameters}

\end{table*}

\begin{figure*}[!htbp]
    \centering
    \includegraphics[scale=0.55]{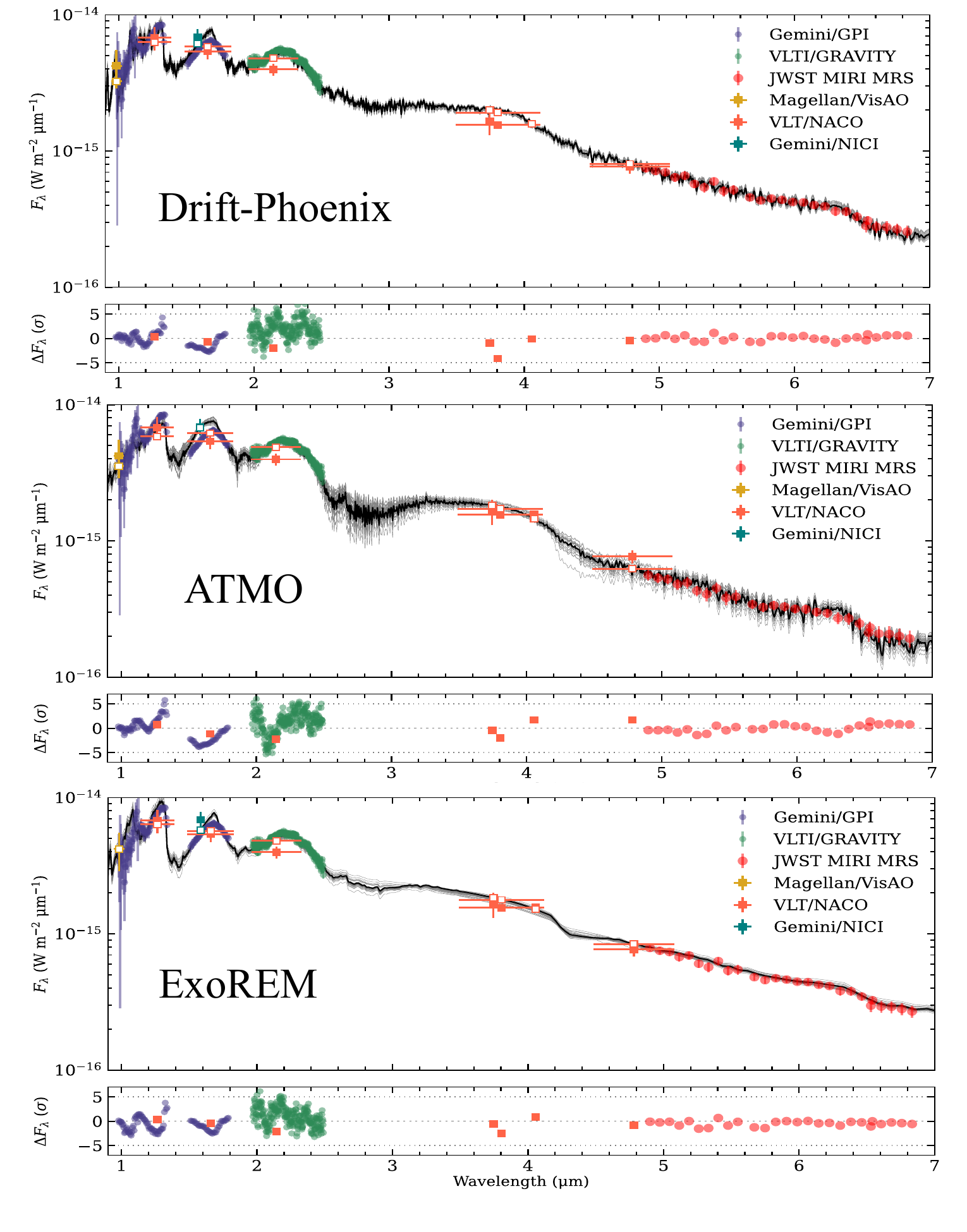}
    \caption{Top: best fit Drift-Phoenix model (black) with spectra and photometry of $\beta$ Pic b. Middle: same as top but now with the best fit model from the ATMO grid. Bottom: same as the top but with the ExoREM grid. The grey spectral models are sampled from the posterior distribution. The residuals from the best fit model are shown in the bottom panel of both plots with the two dotted lines representing a 5$\sigma$ residual. All three models appear to be well fit, but the ExoREM grid yields a systematically lower temperature than the other two. }
    \label{fig:species_fits}
\end{figure*}

 \section{Discussion}

 \subsection{$\beta$ Pic b Atmosphere Composition }

The composition of a giant planet's atmosphere, usually the C/O ratio, can potentially be used as a tracer to infer formation mechanisms and location in the parent protoplanetary disk \citep{Oberg11}. There are two main hypothesis for forming a giant planet. The first is through gravitational collapse, which is similar to star formation \citep{Bodenheimer74}. In this scenario, a region of the protoplanetary disk becomes gravitationally unstable and quickly collapses to form a planet that then cools over time. The second is core accretion where a solid core is formed that slowly accretes gas from the surrounding disk until the mass of the accreted gas is similar to that of the core. Then, a phase of runaway gas accretion occurs and the planet accretes a significant amount of gas over a short period of time \citep{Lissuaer07}. As discussed by the \cite{Nowak20}, in the gravitational collapse scenario, the C/O ratio of the planet is expected to match that of the host star because the formation happens quickly and all solid and gaseous material in the disk that collapses into the planetary atmosphere have a combined stellar C/O ratio. The core accretion scenario takes longer than the gravitational collapse scenario, which provides more time for the accretion of solid materials, including planetesimals. In the core accretion scenario, without enrichment from solids before the runaway gas accretion phase, the atmosphere of the planet is not comprised of the combination of gas and solids like in the gravitational collapse scenario and is only composed of the gas. Therefore, without the enrichment of solids, the C/O ratio of the atmosphere in this scenario is expected to match the C/O ratio of the gas in the disk, which is super-stellar. However, with enrichment from solids before the runaway gas accretion stage, the C/O ratio of the atmosphere can be lowered to sub-stellar values as there is more time to accrete solid material in the core accretion scenario than the gravitational collapse scenario \citep{Nowak20}. 

The wavelength range of our MRS spectrum of $\beta$ Pic b contains a water absorption band, which, along with the CO band in the GRAVITY spectrum of $\beta$ Pic b, could potentially help constrain the C/O ratio of the planet's atmosphere. From our modeling with the ATMO grid of all the spectra and photometry in Figure \ref{fig:species_fits}, we get a C/O ratio for $\beta$ Pic b of 0.39$^{+0.10}_{-0.06}$. This is consistent with the C/O ratio determined by the \cite{Nowak20} from a petitRADTRANS retrieval of the GRAVITY and GPI spectra of $\beta$ Pic b. This retrieval yielded a C/O ratio of 0.43$^{+0.04}_{-0.03}$. The ExoREM model grid gives a  C/O ratio of 0.36$^{+0.13}_{-0.05}$ which also contains the C/O ratio from the \cite{Nowak20} within the uncertainty.

 Using the C/O ratio of a planetary atmosphere to infer formation history requires knowledge of the C/O ratio of the host star, however, there is not a published value for the C/O ratio of $\beta$ Pic in the literature. Because of this, we compared to the C/O ratio measured from the $\beta$ Pic moving group member HD 181327 as in Reggiani et al. (2023, in press). The C/O ratio of HD 181327 was used as a proxy for $\beta$ Pic since they likely formed out of the same molecular cloud. The C/O ratio of HD 181327 is 0.62$\pm$0.08 (Reggiani et al. 2023, in press). The C/O ratio for $\beta$ Pic b we got from the ATMO grid fitting is sub-stellar and does not contain the stellar C/O in its uncertainty. Similarly, the best fit C/O ratio from the ExoREM grid does not contain the stellar C/O. Both C/O ratios from the two model grids are within the uncertainties of each other and suggest a sub-stellar C/O ratio for the planet.

The two C/O ratios we inferred from the grid model fitting are both consistent within the uncertainties with the C/O ratio measured from a free retrieval to the GRAVITY and GPI spectra of 0.43$^{+0.04}_{-0.03}$ \citep{Nowak20}. This sub-stellar C/O ratio was argued to favor the core accretion scenario with planetesimal enrichment to lower the C/O ratio of the planetary atmosphere. Our C/O ratios from the grid fitting are consistent with what was found by the \cite{Nowak20}. The MIRI MRS data and grid model fitting presented, however, is not enough to constrain the C/O ratio, and thus the formation history of $\beta$ Pic b, better than what was presented by the \cite{Nowak20}, so for a complete discussion on the formation history of $\beta$ Pic b, see \cite{Nowak20}.

 \subsection{Dust Accretion onto $\beta$ Pic b and c}\label{sec:Dust_accret}
 The infrared excess seen at 5 $\text{\textmu}$m suggests that there is dust produced in the inner few au of the system. The dust emitting at 5 $\text{\textmu}$m that we see to be spatially extended out to $\sim$20 au in the PSF subtracted image slices is best explained by grains in the small grain limit ($2\pi a\ll\lambda$). This spatially extended dust seen at 5 $\text{\textmu}$m is then likely to be below the blowout size for the system (see Figure \ref{fig:Beta_force} for the blowout size) and can be driven outwards over time by radiation pressure. The blackbody stellocentric distance for the 500 K dust (from our best fit blackbody to the 5 to 7.5 $\text{\textmu}$m excess) is 0.9 au. If small dust grains are being produced at 0.9 au, and are then radiatively driven outwards, some will likely collide with the planets $\beta$ Pic b and $\beta$ Pic c, which have semi-major axes of 9.9$\pm$0.05 au and 2.72$\pm$0.02 au \citep{2Nowak20} respectively. We used our MIRI MRS data to perform an order-of-magnitude estimate for the dust accretion rate from the inner hot dust onto the two known planets in the $\beta$ Pic system.

 To estimate the dust accretion rate onto $\beta$ Pic b and $\beta$ Pic c, we first calculated the dust mass from the 5 $\text{\textmu}$m unresolved excess using the equation from \cite{Lisse09},
 \begin{equation} 
     F_{\lambda}=\frac{1}{D^2}\int_{a_{min}}^{a_{max}}B_{\lambda}(T)Q_{abs}(\lambda)\pi a^2 S\frac{dn}{da}da,
 \end{equation}\label{eqn:Dust_flux_eqn}
where $F_{\lambda}$ is the flux density from the dust at a given wavelength, $D$ is the distance to the star (19.6 pc) \citep{GAIA23}, $B_{\lambda}(T)$ is the blackbody function at a given temperature $T$, $Q_{abs}(\lambda)$ is the absorption coefficient of the material at a given wavelength, $S$ is a scaling factor for the particle size distribution, $a$ is the grain radius, and $\frac{dn}{da}$ is the particle size distribution. We assumed the particle size distribution that is expected for collisional equilibrium where $dn/da \propto a^{-3.5}$ \citep{Dohnanyi69}. We assumed a dust temperature of 500 K based on the fit to the 5 to 7 $\text{\textmu}$m excess and for the absorption coefficient, we calculated $Q_{abs}(\lambda)$ assuming Mie Theory and the optical constants of amorphous olivine (\ce{Mg2SiO4}) from \cite{Jaeger03} for the different dust grain sizes spanning the region $a_{min}=0.03$ $\text{\textmu}$m to $a_{max}=10$ $\text{\textmu}$m. Because we do not know the composition of the dust grains producing the 5 $\text{\textmu}$m excess, we assume they are amorphous silicate grains, since these were detected by Spitzer around $\beta$ Pic \citep{Lu22}. Because of this uncertainty in composition, we also repeat this calculation assuming optical constants of another silicate species, amorphous pyroxene (\ce{MgSiO3}) from \cite{Jaeger94}. 

At 5 $\text{\textmu}$m, the flux density from the excess dust emission is 0.2 Jy (see Figure \ref{fig:blackbody_fit}). We determined a dust mass estimate by scaling the particle size distribution (using the scaling factor $S$) such that the dust flux density calculated on the right hand side of Equation 5 is equal to the observed dust flux density at 5 $\text{\textmu}$m. We then calculated the dust mass by integrating over the scaled particle size distribution, using the equation 
\begin{equation}
    M=S\rho\int_{a_{min}}^{a_{max}} a^{-3.5}\frac{4\pi}{3}a^3da
\end{equation}
 where $M$ is the dust mass, $S$ is the scaling factor from Equation 5, $\rho$ is the bulk density, and $a $ is the grain radius. The $\frac{4\pi}{3}a^3$ term is the volume of a spherical grain and $a^{-3.5}$ is from the assumed particle size distribution. We assumed a bulk density of $\rho=3.3$ g/cm$^{3}$. Integrating Equation 6 gives a dust mass from the 5 $\text{\textmu}$m excess of 7$\times 10^{21}$ g assuming that all of the excess is produced from olivine. Doing the same calculation for pyroxene grains gives a dust mass of 1$\times 10^{22}$ g.

We next calculated the blowout size for olivine and pyroxene silicate grains around $\beta$ Pic. The ratio of the radiation force to the gravitational force can be written as 
\begin{equation}
    \beta=\frac{3L_*\langle Q_{pr}\rangle}{16\pi GM_*ca\rho},
\end{equation}
where $L_*$ is the stellar luminosity, $M_*$ is the mass of the star, $c$ is the speed of light, $G$ is the gravitational constant, $a$ is the radius of the dust grains, $\rho$ is the bulk density of the material, and $\langle Q_{pr}\rangle$ is the average radiation pressure coupling coefficient for a given material. We used a mass of $\beta$ Pic of 1.75 $M_{\odot}$ \citep{Crifo97}. $\langle Q_{pr}\rangle$ is given as 
\begin{equation}
    \langle Q_{pr}\rangle=\frac{\int Q_{pr}F_{\lambda}d\lambda}{\int F_{\lambda}d\lambda}
\end{equation}
where $F_{\lambda}$ is the flux density at a given wavelength of the star. For $F_{\lambda}$, we used the photosphere model from \cite{Lu22} and we calculated $Q_{pr}$ assuming Mie theory and amorphous olivine and pyroxene grains. Dust grains with $\beta$ values $>0.5$ become unbound and are blown outwards, setting the blowout size for the system \citep{Krivov06}. Figure \ref{fig:Beta_force} shows the calculated $\beta$ values as a function of grain size along with a horizontal dashed black line indicating where $\beta=0.5$. Grain sizes with $\beta$ values above the horizontal black line are blown out of the system, which sets the blowout radius for the system to be between 0.03 and 1.14 $\text{\textmu}$m, assuming olivine grains. For pyroxene grains, the blowout radius is between 0.03 and 1.24  $\text{\textmu}$m.

\begin{figure}[!htbp]
    \centering
    \includegraphics[scale=0.55]{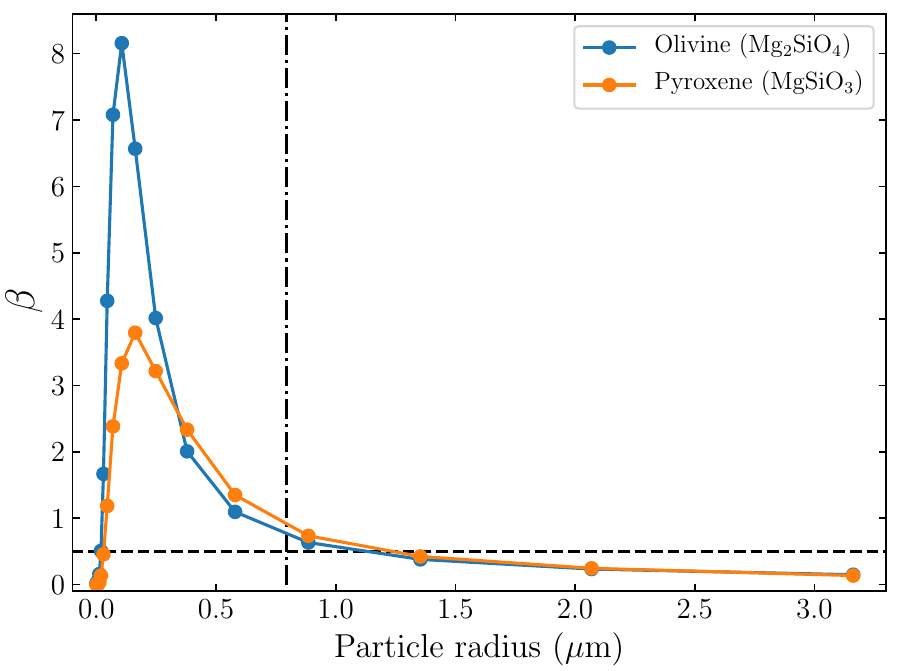}
    \caption{ $\beta$ (ratio of radiation pressure to gravitational force) as a function of particle radius for Olivine (\ce{Mg2SiO4}) and Pyroxene (\ce{MgSiO3}) silicate grains. The black dashed horizontal line shows where $\beta=0.5$. Grains sizes with $\beta>0.5$ become unbound and are blown out of the system. The vertical black dashed dotted line shows $a=\lambda/(2\pi)$ for $\lambda=5$ $\text{\textmu}$m, which is $a=0.8$ $\text{\textmu}$m. The small grain limit is defined as $2\pi a \ll \lambda$.}
    \label{fig:Beta_force}
\end{figure}

Using the $\beta$ values of the dust grains, we calculated their terminal velocity and the time it takes for the dust grains to reach the planets assuming that their initial location is at the blackbody distance of the 5 $\text{\textmu}$m excess (0.9 au) (since we don't know the physical location of the grains producing the unresolved excess). We also assumed that the dust has a distribution that is azimuthally symmetric, and that the dust grains travel at their terminal velocity until they accrete onto the planets. We calculated the terminal radial velocity of the dust grains using this expression from \cite{Su05}:
\begin{equation}
    v_r\simeq [(2GM_*/r_{init})(\beta-1/2)]^{1/2}
\end{equation}
 where $v_r$ is the terminal radial velocity and $r_{init}$ is the stellocentric distance where the dust grains are released (we assumed the minimum blackbody distance of $r_{init}=0.9$ au). 

 To calculate the dust accretion rate onto $\beta$ Pic b and $\beta$ Pic c, we binned the scaled particle size distribution we determined above into radius bins of length 0.1 $\text{\textmu}$m from 0.03 to 1.14 $\text{\textmu}$m (the blowout size range for olivine) and calculated a dust mass in each radius bin by integrating Equation 6 over each radius bin. We then calculated the terminal velocity of the dust grains in each radius bin by calculating $\beta$ for each bin center radius using Equation 7. Assuming that the dust grains start at 0.9 au and using the semi-major axes of $\beta$ Pic b and $\beta$ Pic c of 9.9 and 2.72 au, we then calculated the time it takes for the dust to travel from their initial location to the distance of the planets assuming that they travel the entire distance at the terminal velocity. We checked this assumption by integrating the equation of motion of a dust grain in orbit around $\beta$ Pic in two dimensions assuming that the dust grain starts out on a circular orbit with the Keplerian velocity at 0.9 au. The force in the radial direction for a dust grain under gravitational and radiation forces, as shown in \cite{Lisse98} and \cite{Krivov06}, is 
 \begin{equation}
     \textbf{F}=\frac{-GMm(1-\beta)}{r^3}\textbf{r}
 \end{equation}
 where $r$ is the distance from the dust grain to the central star. From numerically integrating Equation 10 in two dimensions with a $\beta$ value of $\beta$=0.5, we found that the dust grain will reach its terminal velocity in about a tenth of the time it takes to reach the semi-major axis of $\beta$ Pic b and 40$\%$ of the time it takes to get to $\beta$ Pic c. For dust grains of larger $\beta$ values, they reach their terminal velocity quicker. Thus, the assumption that the dust grains travel the entire distance at the terminal velocity likely does not have a large effect on the estimated dust accretion rate.
 
 We estimated the dust accretion rate in each radius bin of the particle size distribution using the equation
 \begin{equation}
     \dot{M}=\frac{M_d\pi R_p^2}{4t\Omega d^2},
 \end{equation}
where $\dot{M}$ is the accretion rate, $M_d$ is the dust mass in each size bin, $R_p$ is the planet radius, $t$ is the time it takes the dust in each radius bin to reach the planet from its origin location, $d$ is the distance between the dust origin radius and the planet semi-major axis, and $\Omega$ is the solid angle subtended by the disk. We used planetary radii of 1.36 $R_J$ for $\beta$ Pic b \citep{Nowak20} and 1.2 $R_J$ for $\beta$ Pic c \citep{2Nowak20}. We summed this equation across all radius bins in the particle size distribution to get a total dust accretion rate.

 To estimate $\Omega$, we first measured the vertical profile of the spatially resolved dust at 5 $\text{\textmu}$m by collapsing the disk along the midplane to create a single integrated vertical profile of the disk with high S/N. Then, we fit the vertical profile with a Lorentzian to measure the FWHM of the disk. The disk vertical structure and the best fit Lorenztian is shown in Figure \ref{fig:Vert_cut}. To measure the FWHM of the disk, we first subtracted in quadrature the FWHM of the PSF from the best-fit FWHM of the Lorentzian. We used the measured FWHM of the PSF described above. The FWHM of the disk measured from the best fit Lorentzian after deconvolving with the PSF FWHM is 11.5$\pm$0.3 au. 

 \begin{figure}[!htbp]
    \centering
    \includegraphics[scale=0.55]{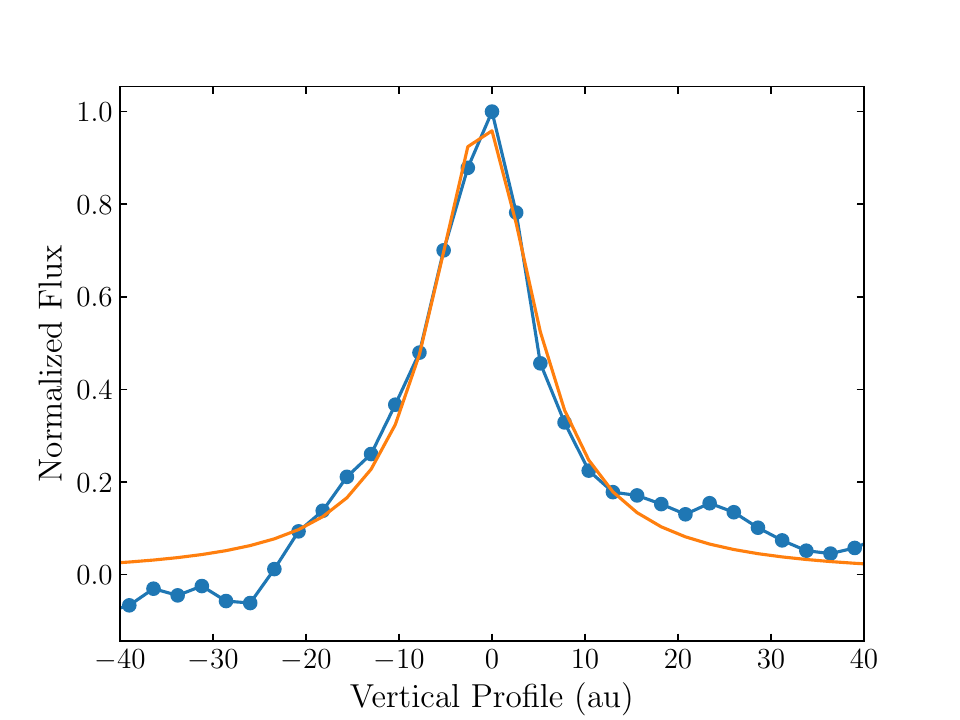}
    \caption{Vertical cut of the disk at 5.2 $\text{\textmu}$m with the best fit Lorentzian profile overlaid.}
    \label{fig:Vert_cut}
\end{figure}

We calculated the solid angle $\Omega$ using the vertical profile of the 5 $\text{\textmu}$m spatially extended disk determined above. We calculated the opening angle for the 5 $\text{\textmu}$m disk by using half the FWHM of the vertical profile (11.5 au) and the outer radius of the disk (20 au) and then solving $tan(\theta)=h/r$ where $h$ is 5.75 au and $r$ is 20 au. This gives an opening angle of $\theta=16^{\circ}$. We then integrated the differential for solid angle $d\Omega$=$\sin(\theta)d\theta d\phi$ over the opening angle and over 2$\pi$ radians azimuthally. This gives a solid angle of $\Omega=1.92$ sr. Then, using Equation 11, we calculated a dust accretion rate for $\beta$ Pic b and $\beta$ Pic c  of $\dot{M}=2\times 10^{-17}$M$_\text{J}$/yr and $\dot{M}=2\times 10^{-15}$M$_\text{J}$/yr assuming Olivine dust grains. Repeating the calculation with Pyroxene dust grains gives dust accretion rates on $\beta$ Pic b and $\beta$ Pic c of $\dot{M}=3\times 10^{-17}$ M$_\text{J}$/yr and $\dot{M}=3\times 10^{-15}$M$_\text{J}$/yr respectively. 

We also performed this calculation for an Fe/Mg ratio of 1:1 for Olivine (\ce{MgFeSiO4}) and Pyroxene (Mg$_{0.5}$Fe$_{0.5}$SiO$_3$), and we found that the $\beta$ values increase for these more iron rich grains compared to the pure magnesium grains. For Pyroxene, using a composition of Mg$_{0.5}$Fe$_{0.5}$SiO$_3$, the $\beta$ values increased by about a factor of 1.4 from those shown in Figure \ref{fig:Beta_force}. The blowout size for Mg$_{0.5}$Fe$_{0.5}$SiO$_3$ is $a\leq$1.8 $\text{\textmu}$m. For this composition of 50$\%$ Fe and Mg, the value for $\beta$ does not come below 0.5 at the small particle size like it does for \ce{MgSiO3} and \ce{Mg2SiO4} as Figure \ref{fig:Beta_force} shows. Instead, all particles below 1.8 $\text{\textmu}$m are blown outwards, so we have to estimate the smallest size grains in the particle size distribution to get a limit on the smallest particle sizes that are present in the disk. \cite{Krijt14} estimated the smallest particle size in a debris disk and relate it to the blowout size with the equation 

\begin{equation}
   \hspace{-0.5cm} \small\frac{s_{min}}{s_{blow}}=2.4\left(\frac{a}{5 \text{au}}\right)\left(\frac{L_*}{L_{\odot}}\right)^{-1}\left(\frac{f}{10^{-2}}\right)\left(\frac{\eta}{10^{-2}}\right)\left(\frac{\gamma}{0.1 \text{J m}^{-2}}\right),
\end{equation}
where $a$ is the semi-major axis of the dust, $L_*$ is the luminosity of the star, $f$ is the ratio of the relative velocity of the colliding bodies and their Keplerian velocity, $\eta$ is the fraction of the pre-collision kinetic energy of the colliding bodies that is converted into making a new surface, and $\gamma$ is the surface energy per unit surface area of the material. As done in \cite{Krijt14}, we assumed that the collisions are below the hyper-velocity regime and $f=10^{-2}$ and $\eta=10^{-2}$. We used $\gamma=0.05$ Jm$^{-2}$ for silicate grains \citep{Krijt14} and we assumed the dust grains are at the blackbody distance inferred from the blackbody fit above of 0.9 au. We find that for silicate grains, the minimum dust particle radius in the disk is 0.028  $\text{\textmu}$m. We then used this size as the minimum particle size for the more iron rich grains (\ce{MgFeSiO4} and \ce{Mg$_{0.5}$Fe$_{0.5}$SiO3}) and calculated the dust accretion rate in the same method as above. We estimated a dust accretion rate onto $\beta$ Pic b for Mg$_{0.5}$Fe$_{0.5}$SiO$_3$ of 6$\times10^{-17}$ M$_\text{J}$/yr and 5$\times10^{-17}$ M$_\text{J}$/yr for \ce{MgFeSiO4}. For $\beta$ Pic c we obtained dust accretion rates of 6$\times10^{-15}$ M$_\text{J}$/yr and 5$\times10^{-15}$ M$_\text{J}$/yr for Mg$_{0.5}$Fe$_{0.5}$SiO$_3$ and \ce{MgFeSiO4} respectively. In this calculation, we did not assume that the grains have any porosity, however, \cite{Arnold19} showed that for silicate grains with 97.5$\%$ porosity around $\beta$ Pic, the blowout size increases to $\sim$10  $\text{\textmu}$m, meaning that more dust particles are blown outwards, increasing the dust accretion rates onto the planets.

Given that the masses of these planets are 9.0$\pm$1.6 $M_J$ for $\beta$ Pic b and 8.2$\pm$0.8 $M_J$ for $\beta$ Pic c \citep{2Nowak20}, this dust accretion rate does not contribute significantly to the total mass of these planets. If, however, the small dust grains that are accreted can remain aloft in the planet's atmosphere, they could potentially impact the observed spectra of the planet at near-IR and mid-IR wavelengths. Based on the near-IR photometry of $\beta$ Pic b, \cite{Bonnefoy13} found that the atmosphere of $\beta$ Pic b is likely dusty, which suggests that there could be an absorption feature in the spectrum of $\beta$ Pic b due to silicates at 10 $\text{\textmu}$m. However, since the thermal emission of the disk at 10 $\text{\textmu}$m dominates over the flux of the planet in our MRS data (see Figure \ref{fig:10_mic_disk}), we were unable to search for silicate absorption in the atmosphere of $\beta$ Pic b. Based on these dust accretion rate estimates, we expect the spectrum of $\beta$ Pic c to be more affected by dust than $\beta$ Pic b, although we cannot test this here because we cannot recover $\beta$ Pic c with the MRS since it is too close to the host star. 

\subsection{[Ar II]}
The detection of circumstellar [Ar II] emission around $\beta$ Pic with the MRS is surprising, so here, we speculate on possible explanations for this detection of argon. Similar to the dust in the disk, the gas around $\beta$ Pic is also subject to blowout from radiation pressure \citep{Beust89}. The radiation pressure for the atomic gas around $\beta$ Pic depends on the number and strength of ground state transitions for a given species (see \citealt{Fernandez06}). \cite{Fernandez06} found that singly ionized argon experiences zero radiation force and is not subject to radiative blowout because it does not have any ground state transitions in the 0.1 to 5 $\text{\textmu}$m range. It is then possible for Ar II to accumulate over time in the $\beta$ Pic disk if it is continuously produced through the destruction of minor bodies in the system. Other atomic species detected around $\beta$ Pic are C, O, Na, Mg, Al, Si, S, Ca, Cr, Mn, Fe, and Ni \citep{Brandeker04,Roberge06}, which are thought to be of secondary origin \citep{Fernandez06}, meaning not from the protoplanetary disk, but rather created from the destruction of minor bodies. Similarly, the molecular CO in the $\beta$ Pic disk is thought to be of secondary origin \citep{Matra17}. 

If the argon is also produced from the destruction of minor bodies, then it could persist in the disk since it does not feel radiation pressure. If we assume that the argon production rate is constant over the age of the system ($\sim$20 Myr), we estimate an argon production rate of 3$\times10^{14}$ kg of argon per year to produce the total argon mass seen with the MRS. For reference, this is roughly equivalent to producing the mass of Halley's Comet (2.2$\times10^{14}$ kg) \citep{Hughes85} of Ar II each year in the $\beta$ Pic system. This estimate is highly uncertain given the uncertainties in the argon mass and the likely incorrect assumption that the gas can persist over millions of years in the disk since it does not feel radiation pressure.

In the solar system, minor bodies such as comets and meteorites have been found to contain argon. The argon found in Solar System meteorites is thought to be primordial presolar gas that has become ``trapped" within the matrices of meteorites \citep{Huss96,Patzer02,Vogel03}. Argon has also been detected in the coma of comets 67P/Churyumov-Gerasimenko \citep{Balsiger15} and Hale-Bopp \citep{stern00}. Argon sublimates at a temperature of 40 K, so the detection of argon in comet 67P/Churyumov-Gerasimenko was interpreted to mean that the comet was formed in a cold outer region of the proto-solar nebula \citep{Balsiger15}. If the minor bodies in the $\beta$ Pic system also contain trapped primordial argon like those in the solar system, it is possible that the argon detected around $\beta$ Pic is from the destruction of minor bodies.

Although the production of argon in the $\beta$ Pic system could potentially be explained by the destruction of minor bodies, the mechanism of ionization is unclear. Using a photoionization code, \cite{Fernandez06} found that argon should remain mostly neutral throughout the $\beta$ Pic disk with an ionization fraction on the order of 10$^{-6}$. Their calculation considered UV photons from the star, the interstellar radiation field (ISRF), and ionization from cosmic rays, and they found that the UV radiation field dominated the ionization. It did not consider, however, the coronal X-ray emission from $\beta$ Pic which was detected by \cite{Hempel04} and \cite{Gunther12}. The ionization fraction of argon predicted by \cite{Fernandez06} would require an unphysically large amount (thousands of Earth masses) of neutral argon in the disk to produce the detected [Ar II] emission seen with the MRS. The ionization of neutral argon could then potentially be due to the X-ray emission from the corona of $\beta$ Pic. However, confirmation of this would require a photoionization model that considers the coronal emission from the star, which is beyond the scope of this paper.

\section{Conclusion}
As a part of JWST GTO program 1294, we present MIRI MRS observations of the $\beta$ Pictoris exoplanetary system. Our main findings are: 

\begin{itemize}
    \item We detect an infrared excess from the unresolved point source from 5 to 7 $\text{\textmu}$m that is best fit by a 500 K blackbody, indicating the presence hot dust in the inner few au of the system.
    
    \item Through PSF subtraction, we detect a spatially resolved dust population emitting at 5 $\text{\textmu}$m extending out to stellocentric distances of $\sim$20 au. This dust population is best explained by small sub-blowout size grains because larger blackbody grains in radiative equilibrium are not hot enough between 10 and 20 au to emit significant radiation at 5 $\text{\textmu}$m.

    \item We detect and spatially resolve circumstellar [Ar II] emission for the first time around $\beta$ Pic. This is also the first argon detection in a debris disk.

    \item The unresolved and resolved hot dust population suggests that dust is produced in the inner few au of the system where the sub-blowout size grains are radiatively driven outward where they could accrete onto $\beta$ Pic b and $\beta$ Pic c. We use our MRS data to estimate a dust accretion rate onto $\beta$ Pic b and $\beta$ Pic c of $\dot{M}= 10^{-17}$ M$_\text{J}$/yr and $\dot{M}= 10^{-15}$ M$_\text{J}$/yr respectively. 

    \item We detect $\beta$ Pic b with both PSF subtraction and cross-correlation with atmosphere models. We find that water is the only molecule we detect in cross-correlation. We also present the first mid-IR spectrum of the planet from $\sim$5 to 7 $\text{\textmu}$m, which includes a water absorption band. 

    \item Grid model fitting for $\beta$ Pic b reveals that the metallicity of the planet is entirely model dependent and that the C/O ratio is likely between 0.3 and 0.5. The atmospheric modelling with both ExoREM and ATMO mode grids appears to favor a sub-stellar C/O ratio for $\beta$ Pic b.

\end{itemize}

KW and CC acknowledge the support from the STScI Director's Research Fund  (DRF) and the NASA FINESST program. This work is supported by the National
Aeronautics and Space Administration under Grant No.
80NSSC22K1752 issued through the Mission Directorate. CL acknowledges support from the National Aeronautics and Space Administration under Grant No. 80NSSC21K1844 issued through the Mission Directorate. I.R. is supported by grant FJC2021-047860-I financed by MCIN/AEI /10.13039/501100011033 and the European Union NextGenerationEU/PRTR. 

\section*{Appendix : Atmosphere Model Fitting Posteriors}
 The resulting corner plots from each of the atmosphere grid model fits to the spectra and photometry of $\beta$ Pic b are shown below. Figure \ref{fig:Phoenix_corner} shows the posterior distributions for the Drift Phoenix grid fit, Figure \ref{fig:Exorem_corner} shows the posterior distributions for the ExoREM grid fit, and Figure \ref{fig:petrus_corner} shows the posterior distributions for the ATMO grid fit.

 \begin{figure*}
    \centering
    \includegraphics[scale=0.47]{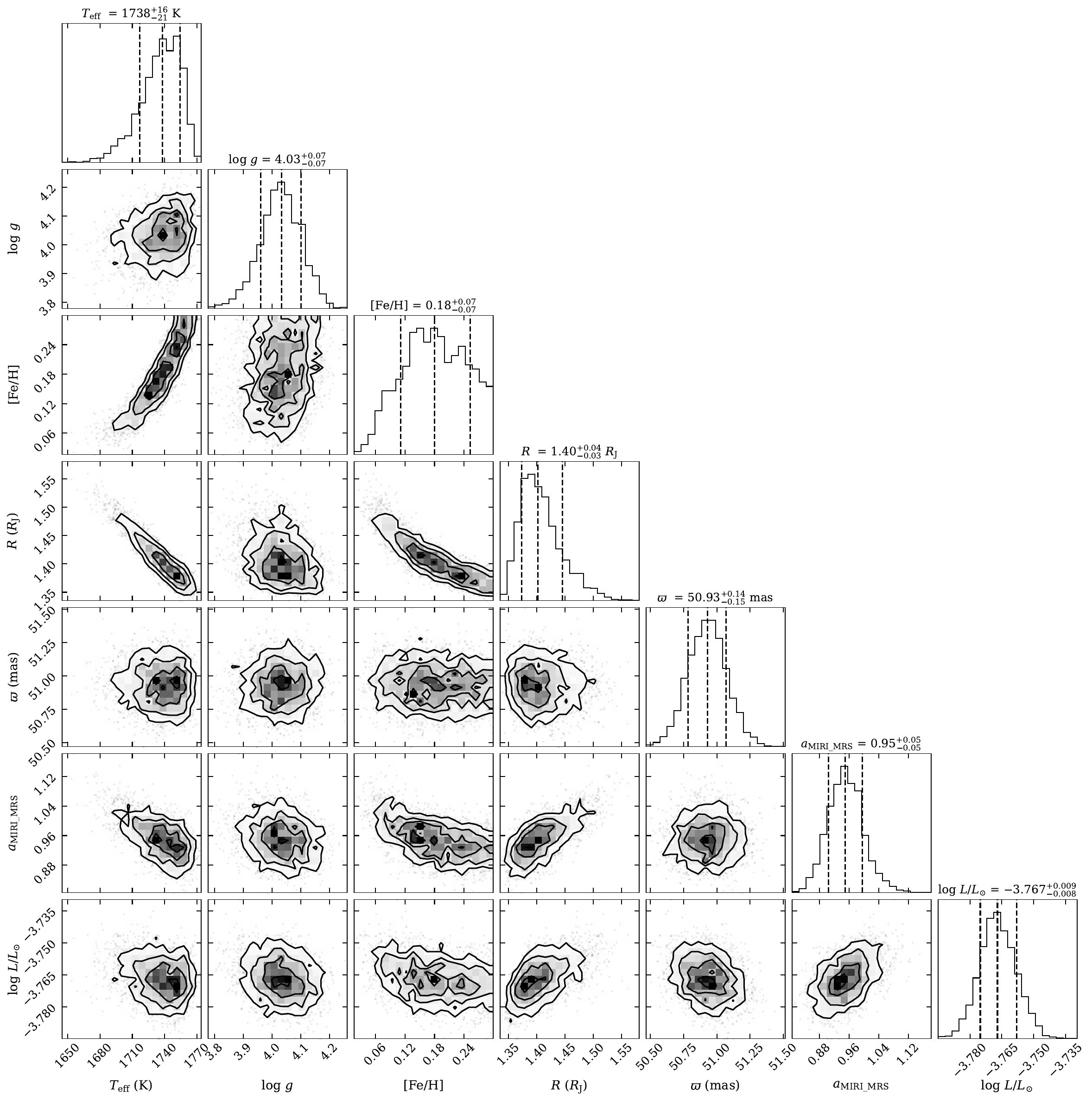}
    \caption{Posterior distribution for the model fitting of the $\beta$ Pic b spectra and photometry of the Drift-Phoenix model grid. The values show the 68$\%$ confidence intervals around the median.}
    \label{fig:Phoenix_corner}
\end{figure*}

 \begin{figure*}
    \centering
    \includegraphics[scale=0.4]{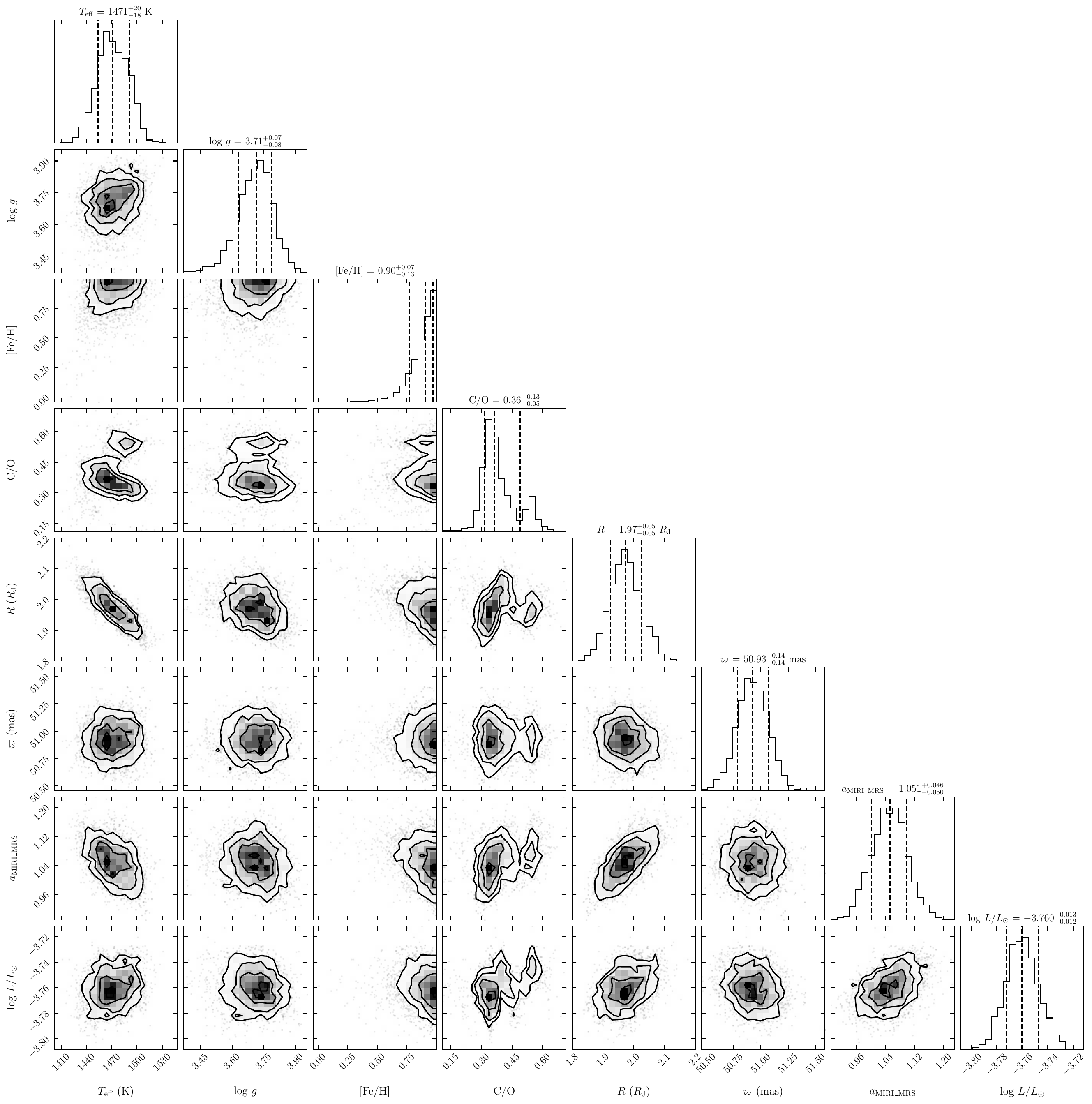}
    \caption{Same as Figure \ref{fig:Phoenix_corner} but for the ExoRem model grid.}
    \label{fig:Exorem_corner}
\end{figure*}

\begin{figure*}
    \centering
    \includegraphics[scale=0.37]{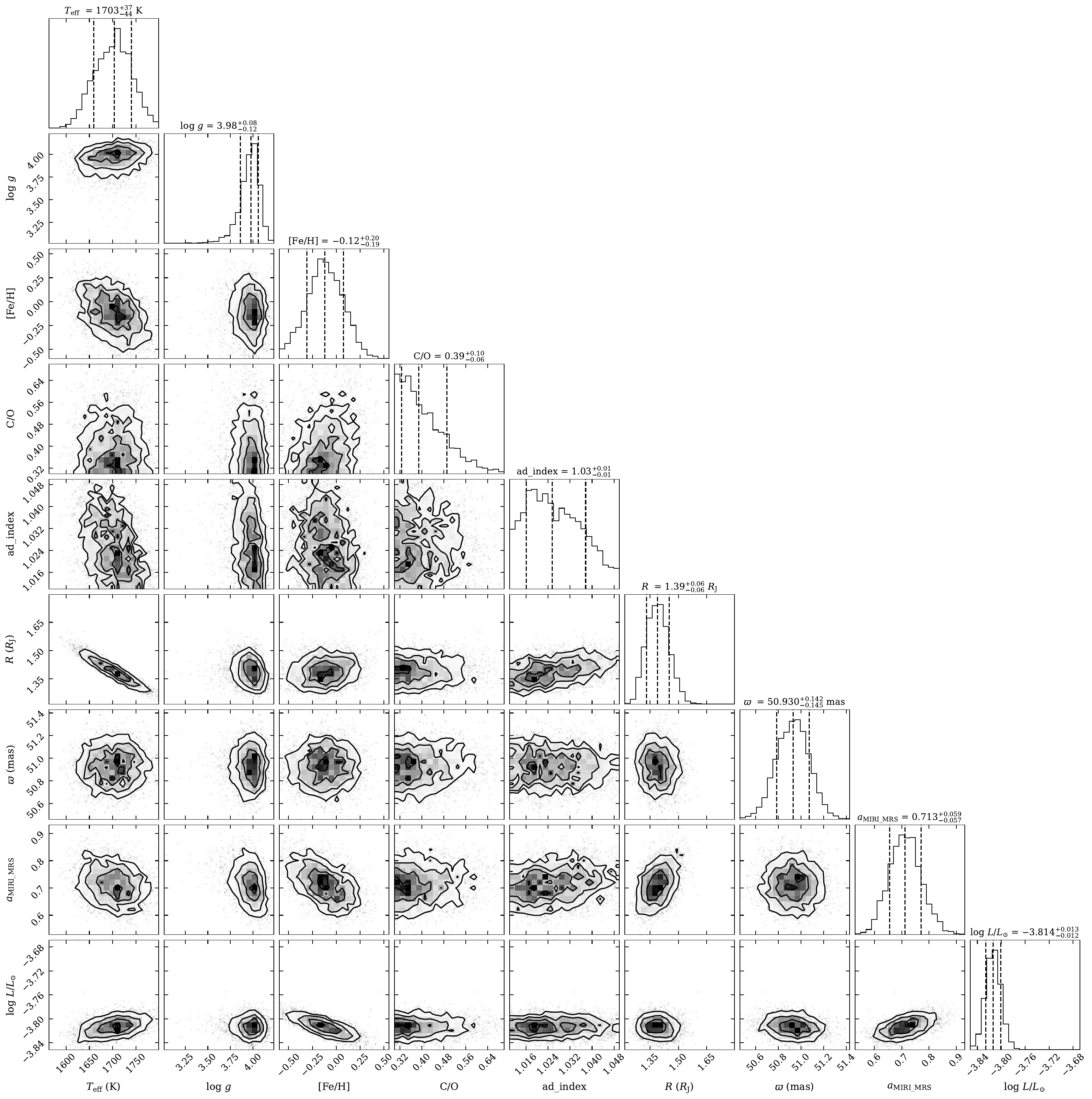}
    \caption{Same as Figure \ref{fig:Phoenix_corner} but for the ATMO model grid from \cite{Petrus23}.}
    \label{fig:petrus_corner}
\end{figure*}

\software{
This research has made use of the following software projects:
    \href{https://astropy.org/}{Astropy} \citep{Astropy13,astropy18,Astropy22},
    \href{https://matplotlib.org/}{Matplotlib} \citep{matplotlib07},
    \href{http://www.numpy.org/}{NumPy} and \href{https://scipy.org/}{SciPy} \citep{numpy07}, Species \citep{Stolker20}, JWST Data Pipeline \citep{Bushouse23}
    and
    the NASA's Astrophysics Data System.
}


\bibliographystyle{aasjournal}
\bibliography{mybib}






\end{document}

%% file: commands.tex
\DeclareSIUnit{\jy}{Jy}
\DeclareSIUnit{\beam}{beam}
\DeclareSIUnit{\kms}{\kilo\meter\per\second}